\documentclass[useAMS,usenatbib]{mn2e}
\usepackage{amsmath}
\usepackage{graphicx}
\graphicspath{{figs/}}

\usepackage{times}

\usepackage{aas_macros} 
\usepackage{url}

\newcommand{\skipthis}[1]{}

\newcommand{\lsim}{${\raisebox{-.9ex}{$\stackrel{\textstyle<}{\sim}$}}$ }
\newcommand{\gsim}{${\raisebox{-.9ex}{$\stackrel{\textstyle>}{\sim}$}}$ }

\def\nh2d{$\rm{NH_2D}$}

\def\nh3{$\rm{NH_3}$}
\def\NH3{$\rm{NH_3}$}

\def\n2hp{$\rm{N_2H^+}$}

\def\h2o{$\rm{H_2O}$}
\def\h2{$\rm{H_2}$}
\def\water{H$_2$O}

\def\meth{CH$_3$OH}

\def\msun{\,$M_\odot$}
\def\Mout{\,$\dot{M}_{\rm out}$}

\def\lsun{\,$L_\odot$}

\def\um{\,$\mu\mathrm{m}$}
\def\mm{\,mm}
\def\kms{\,km~s$^{-1}$}
\def\cm2{\,$\rm{cm^{-2}}$}
\def\cm3{\,$\rm{cm^{-3}}$}
\def\cms{\,$\rm{cm^{-2}}$}
\def\cmc{\,$\rm{cm^{-3}}$}

\newcommand{\ghz}{\,GHz}

\newcommand{\mjy}{\,mJy}
\newcommand{\mjyb}{\,mJy\,beam$^{-1}$}

\def\vlsr{$V\rm{_{LSR}}$}

\def\11{(1,1)}
\def\22{(2,2)}
\def\33{(3,3)}
\def\44{(4,4)}
\def\55{(5,5)}
\def\66{(6,6)}
\def\t21{$T_{21}$}
\def\r31{$R_{31}$}

\newcommand{\trot}{$T_{\rm rot}$}

\newcommand{\tex}{$T_{\rm ex}$}
\newcommand{\tbg}{$T_{\rm bg}$}

\def\ga{G28.34+0.06}
\def\gapa{G28.34-P1}

\def\gb{G11.11{\textendash}0.12}
\def\gbpa{G11.11-P1}
\def\gbpb{G11.11-P6}
\def\gc{G30.88+0.13}
\def\gcca{G30.88-C1}
\def\gccb{G30.88-C2}

\def\alma{Atacama Large Millimeter/submillimeter Array}
\def\atca{Australia Telescope Compact Array}
\def\vla{Very Large Array}

\def\sma{Submillimeter Array}

\def\msx{\emph{Midcourse Space Experiment}}

\def\spt{\emph{Spitzer}}
\def\her{\emph{Herschel}}

\let \ndash = \textendash
\let \amp = \&

\title
[Hierarchical fragmentation in IRDC G11.11--0.12]
{Hierarchical fragmentation and differential star formation in the Galactic ``Snake'': infrared dark cloud \gb}
\author
[Ke Wang et al.]
{\vspace{-1.0cm}
Ke Wang,$^{1,2,3,4,}$\thanks{E-mail: kwang@eso.org}
Qizhou Zhang,$^{2}$
Leonardo Testi,$^{1,5,6}$
Floris van der Tak,$^{3,7}$
\newauthor
Yuefang Wu,$^{4}$
Huawei Zhang,$^{4}$
Thushara Pillai,$^{8}$
Friedrich Wyrowski,$^{9}$
\newauthor
Sean Carey,$^{10}$
Sarah E. Ragan,$^{11}$
and
Thomas Henning$^{11}$\\
\vspace*{6pt} \\
$^1${European Southern Observatory,
Karl-Schwarzschild-Str. 2,
85748 Garching bei M\"{u}nchen,
Germany}\\
$^2${Harvard-Smithsonian Center for Astrophysics, 60 Garden Street,
Cambridge MA 02138, USA}\\
$^3${Kapteyn Astronomical Institute, University of Groningen, Landleven 12, 9747 AD Groningen, The Netherlands}\\
$^4${Department of Astronomy, School of Physics, Peking University,
Beijing 100871, China}\\
$^5$Excellence Cluster Universe, Boltzmannstr. 2, 85748 Garching bei M\"{u}nchen, Germany\\
$^6$INAF -- Osservatorio astrofisico di Arcetri, Largo E. Fermi 5, 50125 Firenze, Italy\\
$^7${SRON Netherlands Institute for Space Research, Landleven 12, 9747 AD Groningen, The Netherlands}\\
$^8${California Institute of Technology, 1200 E California Blvd, Pasadena, CA 91125, USA}\\
$^9${Max-Planck-Institut f\"{u}r Radioastronomie, Auf dem H\"{o}gel 69, D-53121 Bonn, Germany}\\
$^{10}${Spitzer Science Center, California Institute of Technology, Pasadena, CA 91125, USA}\\
$^{11}${Max-Planck Institute f\"{u}r Astronomie, K\"{o}nigstuhl 17, D-69117 Heidelberg, Germany}
\vspace*{-0.5cm}
}

\begin{document}
\date{Accepted 2014 January 16.  Received 2014 January 15; in original form 2013 October 31}
\maketitle

\begin{abstract}
We present SMA $\lambda = 0.88$ and 1.3 mm broad-band observations, and VLA observations in \nh3 $(J,K)=$ \11 up to \55, \water\ and \meth\ maser lines toward the two most massive molecular clumps in IRDC \gb.
Sensitive high-resolution images reveal hierarchical fragmentation in dense molecular gas from the $\sim$1 pc clump scale down to $\sim$0.01 pc condensation scale. At each scale, the mass of the fragments is orders of magnitude larger than the Jeans mass. This is common to all four IRDC clumps we studied, suggesting that turbulence plays an important role in the early stages of clustered star formation.
Masers, shock heated \nh3 gas, and outflows indicate intense ongoing star formation in some cores while no such signatures are found in others. Furthermore, chemical differentiation may reflect the difference in evolutionary stages among these star formation seeds.
We find \nh3 ortho/para ratios of $1.1\pm0.4$, $2.0\pm0.4$, and $3.0\pm0.7$ associated with three outflows, and the ratio tends to increase along the outflows downstream.
Our combined SMA and VLA observations of several IRDC clumps present the most in depth view so far of the early stages prior to the hot core phase, revealing snapshots of physical and chemical properties at various stages along an apparent evolutionary sequence.
\end{abstract}


\section{Introduction} \label{sec:intro}

\begin{figure*}
\centering
\includegraphics[width=0.98\textwidth,angle=0]{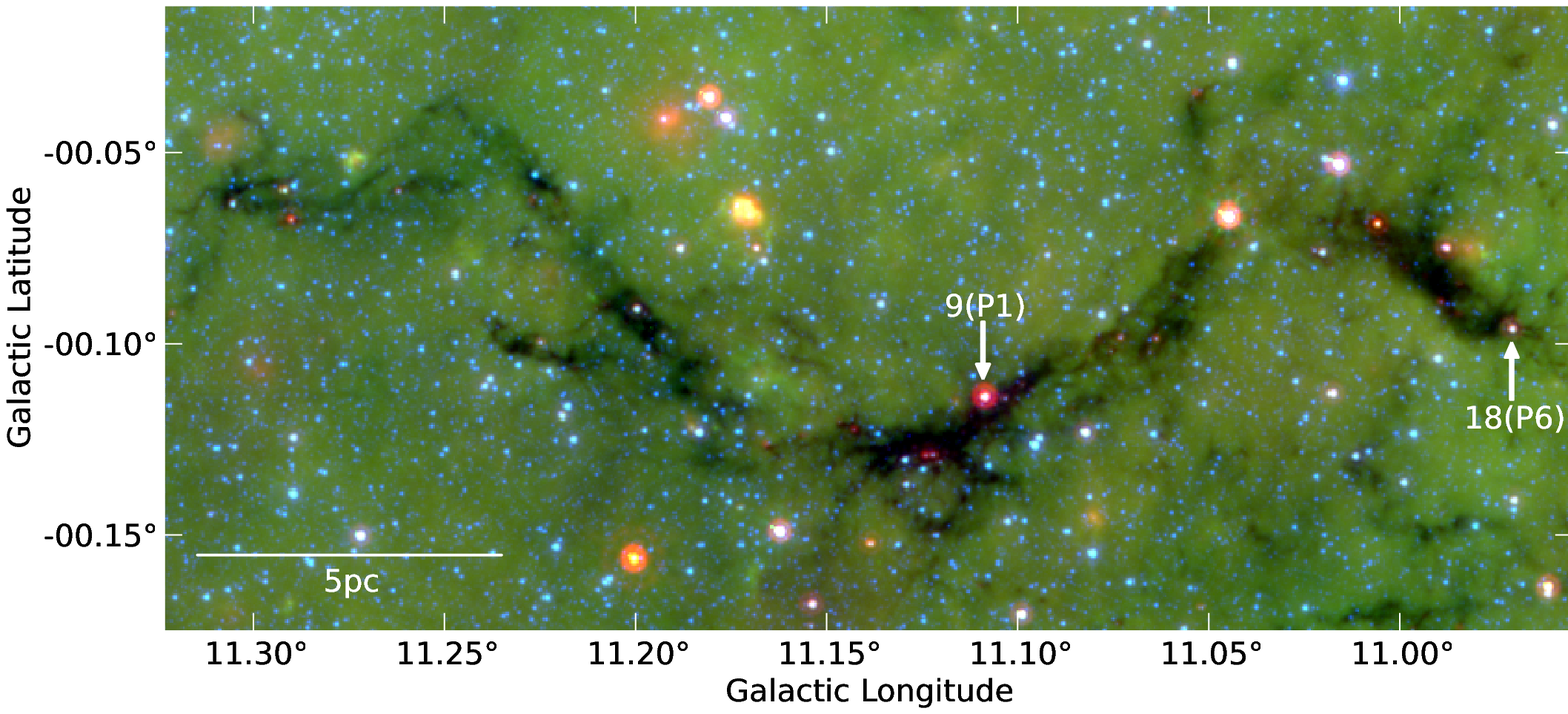}
\caption{
A \emph{Spitzer} composite image (red/green/blue = 24/8/4.5 \um) of the `Snake' nebula.
The \emph{Spitzer} data are taken from the GLIMPSE and MIPSGAL legacy projects \citep{GLIMPSE, MIPSGAL}.
The scale bar indicates the spatial extent of 5 pc at the source distance of 3.6 kpc.
The arrows point to two mid-IR sources, or protostars, as identified by \citet{Henning2010}.
Their ID numbers (9, 18) and associated clumps (P1, P6) are labelled accordingly.
Note this figure is plotted in the Galactic coordinate system, 
while other figures in this paper are in the J2000 Equatorial system.
}
\label{fig:rgb}
\end{figure*}


Because high-mass stars form deeply embedded in dense gas and in distant clustered environments, observational studies face severe limitations of optical depth and spatial resolution. Heavy dust extinction ($N_{\rm H_2}\sim 10^{23} \rm{cm^{-2}}$, $A_{\rm v}>100$ mag) obscures the forming young protostars {and also the Galactic background radiation} even at mid-infrared wavelengths \citep{GLIMPSE,MIPSGAL}, {when viewed against the Galactic plane. This} is what led to the cloud complexes containing such regions to be coined ``infrared-dark clouds'' \citep[IRDCs;][]{perault96,egan98,hennebelle01,simon06a, simon06b,Peretto2009}. The cold dust in IRDCs transitions to emission at longer wavelengths, from far-infrared to sub-millimeter and millimeter wavelengths \citep{Hi-GAL,ATLASGAL,Rosolowsky2010_BGPS2,Aguirre2011_BGPS1}.
One key outcome of the recent \her\ surveys has been the identification of a population of deeply embedded protostellar cores, which appear as point sources ($\sim 0.1$ pc at typical 3 kpc distance to IRDCs) in the \her\ PACS bands \citep{Henning2010,Ragan2012_EPoS}. However, detailed case studies beyond the core scale, the key to understand the early fragmentation that directly initiates subsequent clustered star formation, are only possible by deep interferometric imaging. To maximize the mass sensitivity, the preferred wavelength of observations is in the sub-millimeter regime, accessible by the \sma\ (SMA) and by the recently inaugurated \alma\ (ALMA).

Compared to the numerous interferometric studies on massive proto-clusters \citep[e.g.,][]{rath08,Hennemann2009,Beuther2009,longmore11,Palau2013_1000AU} and \nh3 observations of IRDCs \citep{wy08,Devine11,Ragan2011_NH3,Ragan2012_NH3,me12}, there have been few high angular resolution studies dedicated to pre-cluster clumps in the literature \citep{rath07,Swift2009,qz09,Busquet2010,qz11,me11,Pillai2011,Beuther2013,Lee2013_Orion}.
Among these, only a small portion reached a resolution better than the 0.1 pc core scale. Therefore, in the past years we have used SMA and VLA to peer into several IRDC clumps to study their fragmentation \citep{qz09,qz11,me11,me12,me13_ppvi}. We use SMA  dust continuum emission to resolve hierarchical structures, and VLA \nh3 inversion transitions to precisely measure the gas temperature. We strictly limit our sample to dense molecular clumps that represent the extreme early phases (prior to the appearance of hot molecular cores). This makes our programme unique in probing the early fragmentation.

In one of our studies of \gapa, we found hierarchical fragmentation where turbulent pressure dominates over thermal pressure. This is in contrast with low-mass star formation regions where thermal Jeans fragmentation matches well with observations \citep[e.g.,][]{Lada2008_Pipe}, and is consistent with studies that turbulence becomes more important in high-mass star formation regions \citep[][and references therein]{wy08,me09}.
Whether this kind of fragmentation is a common mode of the initial fragmentation, and how the fragments grow physically and chemically, are of great importance, yet remain unexplored. In this paper, we address these questions by extending our study to two early clumps.
The paper is structured as follows. After a description of the targets (\S\,\ref{sec:targ}) and  observations (\S~\ref{sec:obs}), we present results in \S\,\ref{sec:results} on hierarchical structures (\S\,\ref{sec:str}), masers (\S\,\ref{sec:maser}), outflows (\S\,\ref{sec:outflow}), chemical differentiation of the cores (\S\,\ref{sec:chem}), and \nh3 emission (\S\,\ref{sec:nh3}), followed by discussion in \S\,\ref{sec:discuss} on hierarchical fragmentation (\S\,\ref{sec:frag}), shock enhanced \nh3 ortho/para ratio (\S\,\ref{sec:nh3.RD}), a possible proto-binary with an outflow/disc system (\S\,\ref{sec:p1.disc}), and a global evolutionary sequence of cores and clumps (\S\,\ref{sec:evo.clumps}). Finally, we summarize the main findings in \S~\ref{sec:sum}.

\section{Targets: Dense Clumps in IRDC \gb} \label{sec:targ}
\gb, also known as the ``Snake'' nebula, is one of the first IRDCs identified by \cite{egan98} from the \msx\ images owing to its remarkable sinuous dark features in the mid-IR (see Fig. \ref{fig:rgb} for an overview). 
Shortly after the discovery of \cite{egan98}, \cite{carey98} observed H$_2$CO line emission, a tracer of dense gas, in the central part of the Snake, and thus directly confirmed (in addition to the infrared extinction) the existence of dense gas in the IRDC. A kinematic distance of 3.6 kpc was then inferred based on the radial velocity of the H$_2$CO line, putting the Snake on the near side of the Scutum-Centaurus arm [see \cite{Tackenberg2012} and \cite{Goodman2013AAS}\footnote{See a paper in preparation at \url{https://www.authorea.com/users/23/articles/249}} for a Galactic illustration]. Later, \cite{carey2000} and \cite{Johnstone2003} obtained JCMT 450 and 850\um\ continuum images for the entire Snake, and identified seven major emission clumps P1 through P7. \cite{pillai06, pillai06b} mapped the entire cloud in \nh3\ using the Effelsberg 100-m Radio Telescope and found a consistent \vlsr\ around 29.8\kms\ along the Snake, thus the elongated (aspect ratio 28 pc/0.77 pc = 36:1) cloud is indeed a physically coherent entity, not a chance alignment.

As a demonstration case for the \her\ key project ``Earliest Phases of Star Formation'' (EPoS), \cite{Henning2010} studied \gb\ with deep \her\ images in multiple wavelengths and identified 18 protostellar ``cores'' along the Snake filament, which they call ``seeds of star formation''.  By fitting the spectral energy distributions (SEDs) of individual cores, \cite{Henning2010} obtained physical parameters including dust temperature, mass, and luminosity. Among all these cores, two (\#9 and \#18; see Fig. \ref{fig:rgb} for their locations) massive and luminous cores stand out. With masses of 240 and 82\msun\ and luminosities of $1.3 \times 10^3$ and $1.4 \times 10^2$ \lsun\ respectively, the two cores distinguish from other cores of much lower mass and luminosity, and therefore are the most likely sites to form massive stars in the entire cloud. The two cores reside in clumps P1 and P6 respectively, coincident with two mid-IR point sources which dominate the luminosities of the clumps. P1 and P6 lie in the centre and the head of the Snake respectively. These clumps, with sizes less than 1 pc, are likely results of global fragmentation {(see \S\,\ref{sec:str.cylinder})}, while further fragmentation towards smaller scales are less affected by the global environment but rather depend on local properties of the clumps themselves \citep{Kainulainen2013_Snake}. Both clumps have a mass reservoir of $\sim 10^3$\msun\ within 1 pc (\S\,\ref{sec:frag}). Therefore, P1 and P6 are two massive, relatively low-luminosity molecular clumps that are the most likely sites of high-mass star formation in the ``Snake'' IRDC. Hence, resolving the initial star formation processes in P1 and P6 is of great interest.

Although \gb\ is one of the most well studied IRDCs, previous studies are mostly limited to angular resolution achieved with single dish telescopes \citep{carey98,carey2000,Johnstone2003,pillai06,Tackenberg2012}. The only interferometric studies are still yet to resolve underlying fine structures \citep{pillai06b, gomez2011}. Here, we present new SMA and VLA observations of P1 and P6 which resolve great details of the star formation activities that capture the growth of these star formation seeds in action.

\section{Observations} \label{sec:obs}

\begin{table*}
\centering
\begin{minipage}{180mm}
\caption{Observations \label{tab:obs}}
\begin{tabular}{lllll llccc cc}
\hline
{Telescope} & {Date}      & {Pointing $^a$}  & {Lines} &
\multicolumn{3}{c}{Calibrator$^b$} &  {Bandwidth} &
{Chan. width$^c$}  & {Pol.}  & {Int. Time} \\
\cline{5-7} \\
{Config.}    & {(UT)} & {}  & {}  &
{Gain}  & {Flux}  & {Bandpass}  &{(MHz)}  &
{(\kms)} & {} & {(min)}
\\
\hline
 & \\
\multicolumn{11}{c}{(E)VLA K band (primary beam $2'$)} \\
\cline{4-7}\\
VLA-D    & 2001-Oct-27  & P1.I  & \nh3 \11		& Q1  & Q6  & Q5   &  3.125  &  0.3   &  1  & 10 \\
VLA-D    & 2001-Nov-11  & P1.I  & \nh3 \11		& Q3  & Q6  & Q5   & 3.125  &  0.3   &  1  & 22 \\
VLA-D   & 2004-Aug-24   & P1.II  & \water & Q1  & Q6  & Q1, Q6  & 3.125     &  0.3   &  2  & 20  \\
VLA-D   & 2004-Aug-24   & P1.II  & \nh3 \22, \33 & Q1  & Q6  & Q1, Q6   & 3.125     &  0.6   &  2  & 50  \\
VLA-D   & 2004-Aug-29   & P1.II  & \nh3 \22, \33 & Q1  & Q6  & Q1, Q6   & 3.125     &  0.6   &  2  & 50  \\
EVLA-C   & 2010-Dec-24   & P6.I  & \water, \meth & Q1  & Q6  & Q4   & 4  &  0.8   &  4  & 10,6  \\
EVLA-C   & 2010-Dec-28   & P6.I  & \nh3 \11, \22 & Q1  & Q6  & Q4   & 4  &  0.2   &  2  & 7,8  \\
EVLA-C   & 2011-Jan-18   & P6.I  & \nh3 \22, \33 & Q1  & Q6  & Q4   & 4  &  0.4   &  2  & 20  \\
EVLA-D   & 2013-Mar-17   & P6.I  & \nh3 \11 to \55, \water & Q1  & Q8  & Q7   &   4/8     & 0.1/0.2   &  2  & 45  \\
EVLA-D   & 2013-Apr-18   & P6.I  & \nh3 \11 to \55, \water & Q1  & Q8  & Q7   &   4/8     & 0.1/0.2   &  2  & 45  \\
\hline \\
\multicolumn{11}{c}{SMA 230\ghz\ (1.3\mm) band (primary beam $52''$)}\\
\cline{4-7}\\
SMA-Sub & 2010-Mar-19   & P1.III, P6.II  &Many, see Table \ref{tab:lines}  & Q1, Q2  & M1  & Q4
& $2\times4000$  & 1.1     & 1  & 5.8/6.2 hr $^d$ \\
SMA-Com    & 2010-Jun-15   & P1.III, P6.II  &Many, see Table \ref{tab:lines}  & Q1, Q2  & M1  & Q5, Q7      & $2\times4000$  & 1.1     & 1  &  \\
SMA-Ext     & 2010-Aug-27   & P1.III, P6.II  &Many, see Table \ref{tab:lines}  & Q1, Q2  & M2, M3  & Q7      & $2\times4000$  & 1.1     & 1  &  \\
SMA-Ext     & 2010-Sep-20   & P1.III, P6.II  &Many, see Table \ref{tab:lines}  & Q1, Q2  & M2, M3  & Q7      & $2\times4000$  & 1.1     & 1  &  \\
\hline \\
\multicolumn{11}{c}{SMA 345\ghz\ (880\um) band (primary beam $34''$)}\\
\cline{4-7}\\
SMA-Sub & 2011-Mar-15   & P1.IV, P6.III  &Many, see Table \ref{tab:lines}  & Q1, Q2  & M1  & Q5
& $2\times4000$  & 0.7     & 1  & 2.2/3.5 hr $^d$ \\
SMA-Ext     & 2011-Jul-22   & P1.IV, P6.III  &Many, see Table \ref{tab:lines}  & Q1, Q2  & M3  & Q5      & $2\times4000$  & 0.7     & 1  &  \\
\hline
\end{tabular}

\medskip
\textbf{Note}:
\\$^a${Phase centres in J2000 Equatorial coordinates:
P1.I = 18:10:28.3, -19:22:29;
P1.II = 18:10:30.475, -19:22:29.39;
P1.III = 18:10:28.4, -19:22:38;
P1.IV = 18:10:28.21, -19:22:33.34;
P6.I = 18:10:07.42, -19:29:07.7,
P6.II = 18:10:07.2, -19:28:59;
P6.III = 18:10:07.38, -19:29:08.00.
}
\\$^b${Calibrators are Quasars and Moons:
Q1 = NRAO530 (J1733-130),
Q2 = J1911-201,
Q3 = J1743-038,
Q4 = 3C273,
Q5 = 3C279,
Q6 = 3C286,
Q7 = 3C454.3,
Q8 = 3C48;
M1 = Titan,
M2 = Callisto,
M3 = Ganymede.
}
\\$^c${Native channel width, subject to smoothing for some lines (\S\,\ref{sec:obs.vla}).}
\\$^d${Total on-source integration time combining data from all SMA array configurations for P1 and P6, respectively.}
\end{minipage}
\end{table*}

\subsection{\sma} \label{sec:obs.sma}
\subsubsection{230\ghz\ Band}
The \sma\footnote{The Submillimeter Array 
is a joint project between the Smithsonian Astrophysical Observatory
and the Academia Sinica Institute of Astronomy and Astrophysics and is funded by the
Smithsonian Institution and the Academia Sinica.}
(SMA; \citealt{Ho04}) was pointed towards \gb\ P1 and P6 to obtain
continuum and spectral line emission in the 230 GHz band 
during four tracks in 2010,
when SMA was in its subcompact, compact, and extended configurations.
Time-dependent antenna gains were monitored by periodic observations 
of quasars NRAO530 and J1911-201;
frequency-dependent bandpass responses were calibrated by 
quasars 3C273, 3C279 and 3C454.3;
and absolute flux was scaled by observed correlator counts 
with modelled fluxes of Solar system moons Titan, Callisto, and Ganymede.
The empirical flux uncertainty is about 15\%.
For the four tracks,
we used the same correlator setup 
which covers 4 GHz in each of the lower and upper sidebands (LSB, USB),
with a uniform channel width of 0.812\,MHz (equivalent velocity 1.1 \kms\ at 230\ghz) across the entire band.
System temperatures varied from 80 to 150\,K,
and the zenith opacity at 225 GHz ranges from 0.05 to 0.12
during the four tracks.
The full width at half-maximum (FWHM) primary beam is 
about 52$''$ at the observed frequencies.
Table \ref{tab:obs} summarizes the observations.

The visibility data were calibrated using the IDL superset 
MIR\footnote{\url{ http://www.cfa.harvard.edu/~cqi/mircook.html}}.
Calibrated visibility data were then exported out for imaging and analysis
in \textsc{MIRIAD}\footnote{\url{ http://www.cfa.harvard.edu/sma/miriad,
http://www.astro.umd.edu/~teuben/miriad
}} \citep{Sault1995_Miriad}
and \textsc{CASA}\footnote{\url{ http://casa.nrao.edu}} \citep{Petry2012_CASA}.
Data from different tracks were calibrated separately,
and then combined in the visibility domain for imaging.
Continuum emission was generated by averaging line free channels in the visibility domain.
Table \ref{tab:img} lists the synthesized beam and $1\sigma$ RMS noise of the images.

\subsubsection{345\ghz\ Band}

In 2011, we revisited P1 and P6 with SMA at the 345\ghz\ band in two tracks, one in subcompact and another in extended array configuration. The two tracks used the same correlator setup which covers rest frequencies 333.7--337.7\,GHz in the LSB, and 345.6--349.6\,GHz in the USB, with a uniform spectral resolution of 0.812\,MHz (or 0.7 \kms) across the entire band. System temperatures were in the range of 200--300 K, and the zenith opacity $\tau_{\rm 225\,GHz}$ was stable at 0.06 during the observations. Other parameters are listed in Table \ref{tab:obs}.
The data were reduced and imaged in the same way as the 230\ghz\ data.
Additionally for P1, we made an image using data from the extended configuration only and achieved a higher resolution (see \S~\ref{sec:outflow.p1}).
Image properties are tabulated in Table \ref{tab:img}.

\begin{table*}
\begin{minipage}{90mm}
\caption{Image Properties \label{tab:img}}
\begin{tabular}{lccccc}
\hline
  {}
& \multicolumn{2}{c}{P1}
& {}
& \multicolumn{2}{c}{P6}
\\
\cline{2-3} \cline{5-6}
\\
  {Image$^a$} 
& {Beam} 
& {RMS$^b$} 
& {} 
& {Beam} 
& {RMS$^b$}
\\
\hline
1.3\mm\ continuum
& $2''.2 \times 1''.6$	& 0.9&
& $2''.1 \times 1''.5$	& 0.9\\
1.3\mm\ spec. lines $^c$
& $2''.7 \times 1''.7$	& 25--30&
& \multicolumn{2}{c}{same as P1}\\
880\um\ continuum
& $1''.6 \times 1''.2$	& 3.3&
& $1''.2 \times 1''.0$	& 2.3\\
&&&
& $0''.8 \times 0''.6$	& 1.7\\
880\um\ spec. lines $^a$
& $2''.1 \times 1''.3$	& 110&
& \multicolumn{2}{c}{same as P1}\\
\nh3 \11
& $5'' \times 3''$	& 14&
& $7''.1 \times 2''.8$	& 4\\
\nh3 \22
& $5'' \times 3''$	& 3.5&
& $6''.1 \times 2''.7$	& 2.5\\
\nh3 \33
& $5'' \times 3''$	& 6.5&
& $6''.9 \times 2''.9$	& 2.8\\
\nh3 \44
&\multicolumn{2}{c}{No data}&
& $7''.7 \times 3''.0$	& 2.3\\
\nh3 \55
&\multicolumn{2}{c}{No data}&
& $6''.7 \times 3''.0$	& 2.5\\
\water\ maser
& $5'' \times 3''$	& 13&
& $2''.4 \times 1''.0$	& 2.7\\
&&&
& $7''.0 \times 3''.2$	& 4\\
&&&
& $7''.2 \times 3''.2$	& 2\\
\meth\ maser class I
&\multicolumn{2}{c}{No data}&
& $2''.0 \times 0''.9$	& 2.2
\\
\hline
\end{tabular}

\medskip
\textbf{Note}:
\\$^a${All images are made with natural weighting to achieve the highest sensitivity.}
\\$^b${$1\sigma$ RMS noise in \mjyb.}
\\$^c${For spectral line images, beam varies slightly from line to line, hence a typical beam is listed.}
\end{minipage}
\end{table*}

\subsection{\vla} \label{sec:obs.vla}

The Karl G. Jansky \vla\ (VLA) of NRAO\footnote{The National Radio Astronomy Observatory
is a facility of the 
National Science Foundation operated under cooperative agreement by Associated Universities, Inc.}
was pointed towards P1 in its D configuration in two observation runs in 2001 to observe the \nh3 $(J,K)=(1,1)$ transition
(project AW571, PI: Friedrich Wyrowski).
The 3.125 MHz band was split into 128 channels
with a channel spacing of 24.4 kHz (or 0.3 \kms).
In 2004, P1 was revisited with another phase centre
to obtain 22 GHz \water\ maser and \nh3 \22 and \33 transitions (project AP475, PI: Thushara Pillai).
The \water\ maser was observed in a single IF, dual polarization mode,
with the same bandwidth and channel spacing as the 2001 observations.
The \nh3 \22 and \33 were observed in a 2IF, dual polarization mode,
splitting the 3.125 MHz band into 64 channels, each with a 0.6 \kms\ channel width.

Observations of P6 were carried out using the expanded VLA, or EVLA, in its C configuration in three observation runs during the EVLA early science phase in 2010--2011, and two runs in the 2013 D configuration. Thanks to the flexibility of the new correlator, we observed the \nh3 \11, \22, \33, \44, and \55 transitions as well as 22 GHz \water\ maser and 25 GHz class I \meth\ maser lines, with various bandwidth and spectral resolutions (see Table \ref{tab:obs}).

\begin{figure*}
\centering
\includegraphics[height=0.9\textwidth,angle=-90]{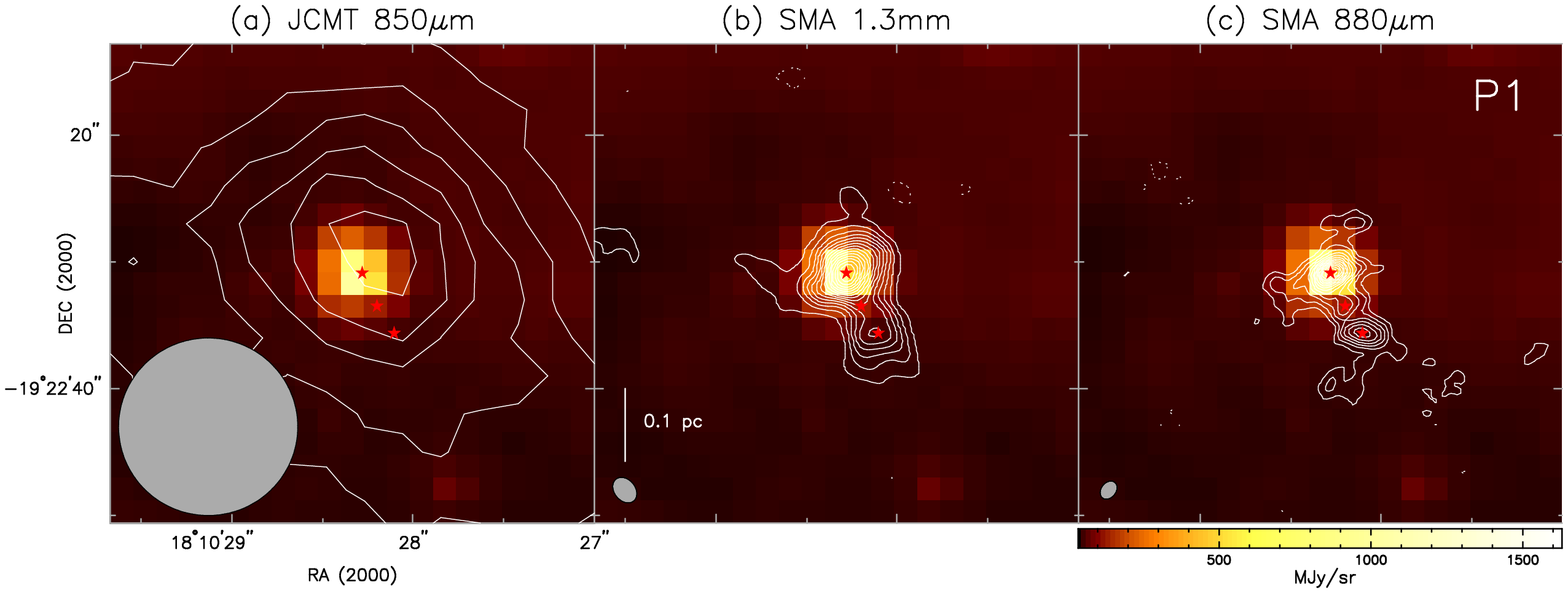}
\vskip 0.1in
\includegraphics[height=0.9\textwidth,angle=-90]{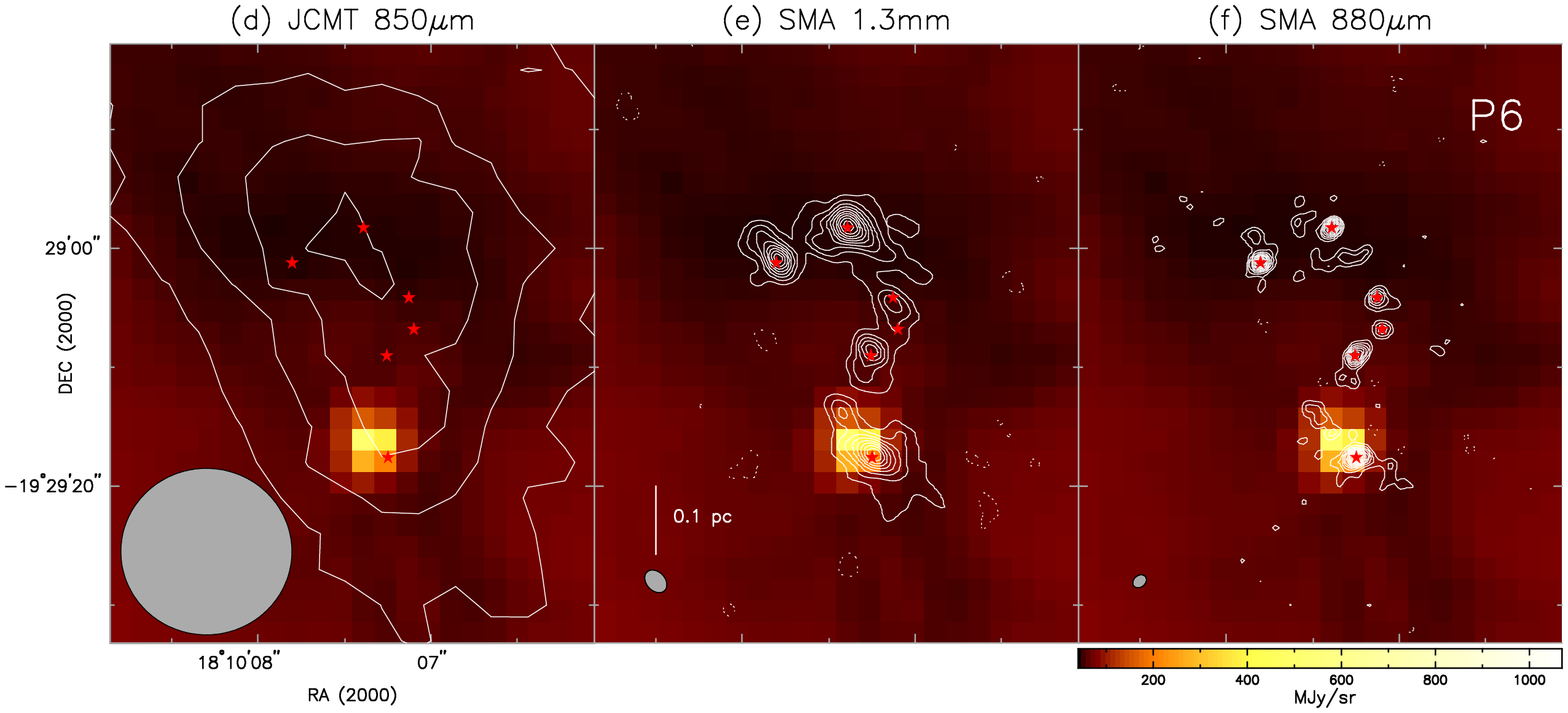}
\caption{
Hierarchical structures in G11.11-P1 (a--c) and G11.11-P6 (d--f),
as seen by JCMT at 850 \um, SMA at 1.3 mm, and SMA at 880 \um\ (contours),
superposed on \emph{Spitzer} 8 \um\ image (color scale).
The JCMT continuum emission is contoured at 0.3 Jy beam$^{-1}$.
The SMA 1.3 mm emission is contoured at
$\pm (3,6,9,...) \sigma$, where $\sigma = 0.9$ mJy beam$^{-1}$
for both P1 and P6.
The SMA 880 \um\ emission is contoured at
$\pm (3,5,7,...) \sigma$, where $\sigma = 3.3$ mJy beam$^{-1}$
for P1 and $\sigma = 2.3$ mJy beam$^{-1}$ for P6.
The shaded ellipse in the bottom left corner of each panel represents 
the beam size of the contoured image.
The red stars mark dominant condensations identified from the SMA 880 \um\ images.
The scale bars represent a spatial scale of 0.1 pc at the source distance of 3.6 kpc.
Negative contours are dashed throughout this paper.
}
\label{fig:con}
\end{figure*}

For all the experiments,
gain, bandpass, and flux variations were calibrated by strong, 
point-like quasars. See Table \ref{tab:obs} for details.
The visibility data were calibrated using \textsc{CASA},
and then imaged and analyzed in \textsc{MIRIAD} and \textsc{CASA}.
Data from different observation runs were calibrated separately, and then combined in the visibility domain for imaging with the exception for maser for which we make individual images from different observing runs, to investigate potential time variations.
The final image cubes keep the native channel width listed in Table \ref{tab:obs}, except for the \nh3 \11, \22, and \33 images of P6 where the final channel width is 0.4 \kms\ in order to include both the C and D configuration data.
Observational parameters are summarized in Table \ref{tab:obs} and image properties are listed in Table \ref{tab:img}.

\section{Results} \label{sec:results}

\subsection{Hierarchical Structure} \label{sec:str}

In the literature, there have been different definitions for a clump, core, and condensation when describing the spatial structure of molecular clouds. For consistency, we adopt the terminology of \textit{clump}, \textit{core}, and \textit{condensation} suggested by \cite{qz09} and \cite{me11}. We refer 
a {clump} as a structure with a size of $\sim 1$ pc,
a {core} as a structure with a size of $\sim 0.1$ pc,
and a {condensation} as a substructure of $\sim 0.01$ pc within a core.
A clump is capable of forming a cluster of stars,
a core may form one or a small group of stars,
and a condensation can typically form a single star or a multiple-star system.
These structures are dense enough that gas and dust are well coupled \citep{Goldsmith2001_Tdust}, thus we use the \nh3 gas temperature as a direct measure of the dust temperature throughout this paper.

\begin{table*}
\centering
\begin{minipage}{110mm}
\caption{Physical Parameters of the Condensations \label{tab:cond}}
\begin{tabular}{lcc rcr ccr}
\hline
{Source}
&{RA}
&{Dec}
&{Flux $^a$}
&{$T$}
&{Mass $^b$}
&\multicolumn{3}{c}{{Size $^c$}}\\
\cline{7-9} \\
{{ID}}
&{(J2000)}
&{(J2000)}
&{(mJy)}
&{(K)}
&{(\msun)}
&{Maj. $('')$}
&{Min. $('')$}
&{PA $(^{\circ})$}\\
\hline
P1-SMA\\
\hline
1	&18:10:28.27 	&-19:22:30.9 	&205.0 &25	&14.1 	&2.1 	&1.3 	&114 \\
2	&18:10:28.19 	&-19:22:33.2 	&46.7  &19	&4.7  	&1.9 	&0.4 	&86  \\
3	&18:10:28.09 	&-19:22:35.7 	&197.0 &17	&23.7 	&3.4 	&1.7 	&69  \\
4	&18:10:28.15 	&-19:22:27.0 	&57.6  &18	&6.3  	&2.5 	&1.2 	&99  \\
5	&18:10:28.56 	&-19:22:31.9 	&53.5  &18	&5.9  	&2.1 	&1.1 	&60  \\
6	&18:10:28.27 	&-19:22:39.7 	&54.5  &15	&8.0  	&2.3 	&1.6 	&156 \\
\hline
P6-SMA\\
\hline
 1 	&18:10:07.80	&-19:29:01.2	&109.7 	&11	&27.9  	&1.1 	&0.9 	&147 \\
1b 	&18:10:07.83	&-19:28:59.3	&83.8 	&11	&21.3  	&3.4 	&1.3 	&48  \\
 2 	&18:10:07.39	&-19:28:58.3	&78.8 	&10	&24.2  	&1.5 	&1.1 	&135 \\
2b 	&18:10:07.58	&-19:28:57.9	&33.0 	&10	&10.1  	&1.5 	&0.7 	&167 \\
2c 	&18:10:07.22	&-19:29:00.6	&38.2 	&10	&11.7  	&2.1 	&0.6 	&131 \\
2d 	&18:10:07.30	&-19:29:00.8	&54.9 	&10	&16.9  	&2.5 	&1.0 	&91  \\
2e 	&18:10:07.39	&-19:29:01.3	&35.2 	&10	&10.8  	&2.5 	&0.7 	&109 \\
 3 	&18:10:07.12	&-19:29:04.3	&66.4 	&14	&10.9  	&1.7 	&1.2 	&55  \\
 4 	&18:10:07.10	&-19:29:06.8	&35.3 	&15	&5.2   	&1.0 	&1.0 	&83  \\
 5 	&18:10:07.25	&-19:29:09.0	&82.1 	&21	&7.2   	&1.6 	&0.8 	&143 \\
5b 	&18:10:07.32	&-19:29:10.8	&30.9 	&21	&2.7   	&1.4 	&1.2 	&16  \\
5c 	&18:10:07.27	&-19:29:11.0	&21.5 	&21	&1.9   	&1.7 	&1.1 	&23  \\
 6 	&18:10:07.25	&-19:29:17.6	&157.0 	&19	&15.9  	&1.4 	&1.0 	&105 \\
6b 	&18:10:07.51	&-19:29:13.5	&29.0 	&19	&2.9   	&1.3 	&0.7 	&50  \\
6c 	&18:10:07.46	&-19:29:14.3	&44.5 	&19	&4.5   	&2.0 	&0.9 	&41  \\
6d 	&18:10:07.39	&-19:29:15.4	&28.7 	&19	&2.9   	&1.3 	&1.0 	&35  \\
6e 	&18:10:07.06	&-19:29:18.4	&32.7 	&19	&3.3   	&1.5 	&0.6 	&36  \\
\hline
\end{tabular}

\medskip
\textbf{Note}:
\\$^a${ Integrated flux obtained from 2D Gaussian fitting and corrected for primary beam attenuation.}
\\$^b${ Mass computed assuming dust opacity index $\beta = 1.5$. The mass scales with $\beta$ in a form of $M \propto 3.5^{\beta}$. For reference, if $\beta = 2$, the mass will be 1.87 times larger.}
\\$^c${ Deconvolved source size.}
\end{minipage}
\end{table*}

Dust continuum images at various resolutions reveal hierarchical structures in P1 and P6. Figure \ref{fig:con} plots JCMT 850\um, SMA 1.3\mm, and SMA 880\um\ images superposed on \spt\ 8\um\ images. The JCMT images outline IRDC clumps P1 and P6, where P1 exhibits a bright MIR source (protostar \#9 in Fig. \ref{fig:rgb}) in its centre, while P6 shows a less bright MIR source (protostar \#18 in Fig. \ref{fig:rgb}) in its southern part. 
As the resolution increases, structures at different scales are highlighted: from $\sim$1 pc scale clumps seen in JCMT 850\um\ images, to the $\sim$0.1 pc scale cores resolved by the SMA 1.3\mm\ images, and to the $\sim$0.01 pc scale condensations resolved by the SMA 880\um\ images.
These structures show in general a good spatial correlation with the \her\ 70\um\ emission and the \spt\ 8/24\um\ extinction, where two IR-bright protostars have already developed (Fig. \ref{fig:con}; \citealt{Henning2010,Ragan2012_EPoS}).

We identify the smallest structure, condensations, based on the highest resolution SMA 880\um\ images as shown in Fig. \ref{fig:con}(c,f). 
All features above $5\sigma$ rms are identified as in \cite{me11} and \cite{qz09}.
We first identify ``major'' emission peaks with fluxes $>9\sigma$ (the 4th contour) and assign them as SMA1, SMA2, SMA3,..., in order from east to west and from north to south. Three major peaks are identified in P1 and six identified in P6, denoted by red stars in Fig. \ref{fig:con}. Then we identify ``minor'' emission peaks with fluxes $>5\sigma$ (the 2nd contour). Three minor peaks are identified in P1, and we assign them as SMA4, SMA5, SMA6, from north to south. In P6, 11 minor peaks are identified. These emission peaks are relatively weak and are associated in position with the 6 major peaks. We thus assign these minor peaks to the associated major peaks. For instance, the two minor peaks associated with P6-SMA5 are assigned as P6-SMA5b and P6-SMA5c. All the identified major and minor peaks are of the size of condensations. Associated condensations may have been fragmented from a common parent core. In summary, we identify 6 condensations in P1 which may belong to 6 cores (P1-SMA1,2,3,4,5,6), respectively, and 17 condensations in P6 which may belong to 6 cores (P6-SMA1,2,3,4,5,6), respectively. For each condensation, we fit a 2D Gaussian function to the observed SMA 880\um\ image and list the results in Table \ref{tab:cond}. All except one (P6-SMA4) ``major'' condensations (red stars in Fig. \ref{fig:con}) coincide with the cores resolved in the SMA 1.3\mm\ images. Condensation P1-SMA1 is coincident with protostar \#9 identified by \cite{Henning2010} from multi-band \her\ images, and condensation P6-SMA6 is coincident with the protostar \#18.

Dust mass is estimated with the assumption that dust emission is optically thin (which is valid at 0.88 and 1.3\,mm), following
\begin{equation}
M_{\rm dust}=\frac{S_{\nu}d^2}{B_{\nu}(T_{\rm dust}){\kappa}_{\nu}}\,,
\label{eq:dust-mass}
\end{equation}
where $M_{\rm dust}$ is the dust mass, $S_{\nu}$ is the continuum flux at
frequency $\nu$, $d$ is the source distance, $B_{\nu}(T_{\rm dust})$ is the
Planck function at dust temperature $T_{\rm dust}$, and
${\kappa}_{\nu}=10({\nu}/1.2\,{\mathrm {THz}})^{\beta}$
cm$^2$\,g$^{-1}$ is the dust opacity \citep{hildebrand83}.
In the calculation we adopt the temperature measured from \nh3 (\S~\ref{sec:nh3.temp}) and the dust opacity index $\beta =1.5$. If $\beta =2$ the mass would be a factor of 2 larger. Dust mass is then translated to gas mass accounting for a gas:dust ratio of 100. The computed total mass for each condensation is reported in Table \ref{tab:cond}.
Note that interferometric images filter out relatively smooth emission due to imperfect $(u,v)$ sampling, leading to ``missing flux''. Thus the total mass of the dense cores or condensations revealed by SMA (Fig.~\ref{fig:con}) is less than the clump mass determined from single dish JCMT observations. Our analysis (\S~\ref{sec:frag}) does not rely on the smooth emission but on clumpy structures, therefore is not affected by the missing flux.
{For reference, the SMA 880\um\ images recover 7\% and 14\% of the total JCMT 850\um\ flux in P1 and P6 , respectively. This is consistent with the fact that P6 contains more compact structures than P1 (Fig.~\ref{fig:con}).}

The sensitivity in the 880\um\ images (Figure \ref{fig:con}, Table \ref{tab:img}) corresponds to 0.2--0.5\msun\ for the 15--25 K gas temperature in the P1 clump (\S~\ref{sec:nh3.temp}), and 0.2--0.7\msun\ in P6 for a 10--21 K temperature range (\S~\ref{sec:nh3.temp}). Therefore, the identified condensation is complete to a $5\sigma$ mass limit of 1--3.5\msun, depending on the temperature.

\begin{figure*}
\centering
\includegraphics[width=0.8\textwidth,angle=0]{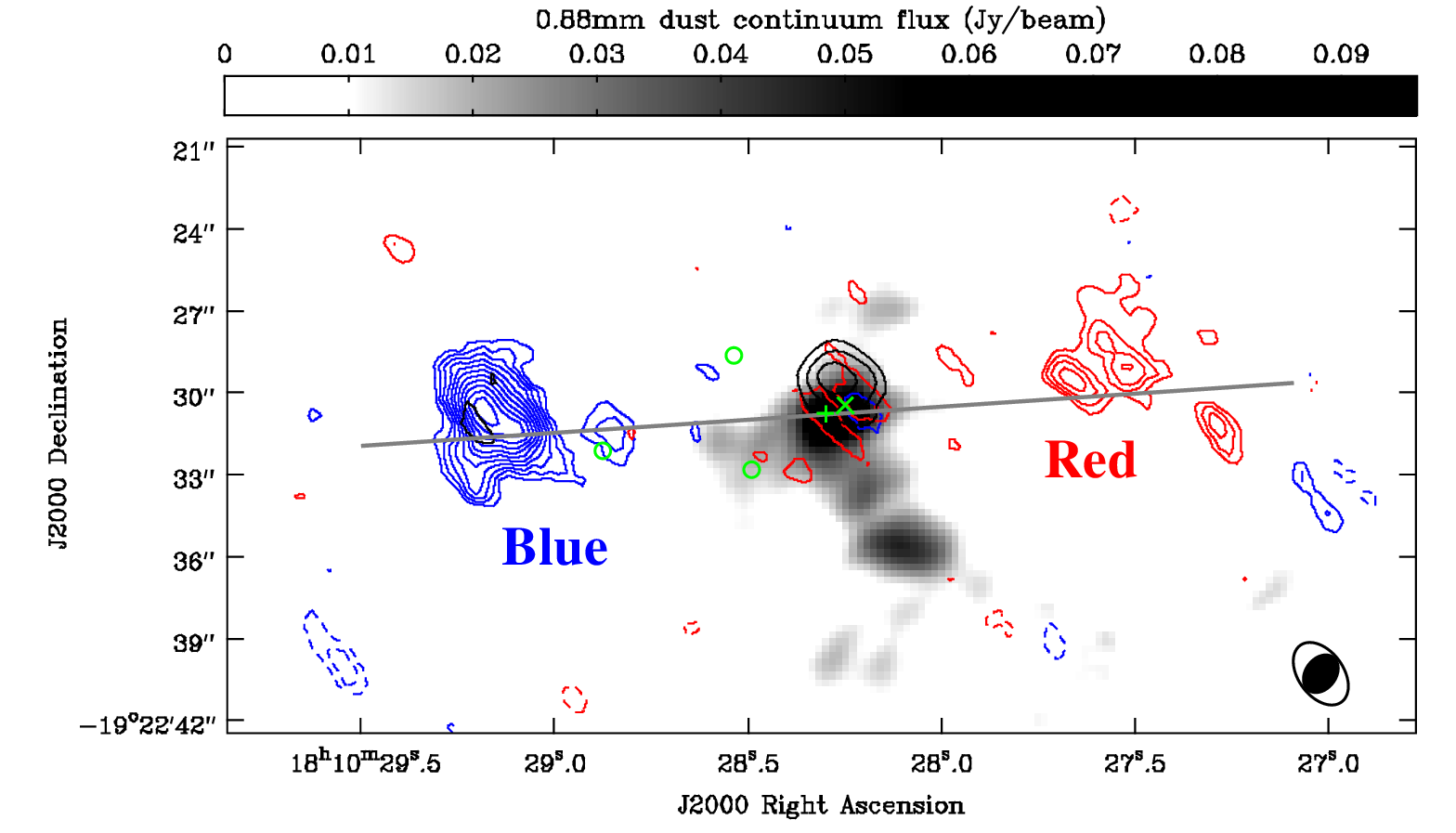}
\caption{
The SiO(5--4) outflow in \gbpa.
Blue/red contours show the blue-/red-shifted SiO emission integrated over [18, 28]\kms\ and [32, 39]\kms, respectively, whereas the {black contours} represent SiO emission near the systemic velocity integrated over [29, 31]\kms. The SiO contours are $\pm(3,4,5,...)\sigma$, where $\sigma = 85$\mjy\,\kms. For comparison, the SMA 880\um\ continuum image is shown {as grey scale in the background}.
Labelled symbols denote \water\ maser W2 ($+$), class II \meth\ maser M1 ($\times$), and the three 2MASS point sources ($\circ$).
In the bottom {right} corner the {open/filled} ellipses represent synthesized beams of the SiO/880\um\ images.
The thick grey line marks schematically the underlying bipolar jet that propels the observed molecular outflow. The grey line centres on W2 and extends 0.3 pc eastbound and westbound, respectively.
}
\label{fig:p1.outflow}
\end{figure*}

\subsection{\water\ and \meth\ Masers} \label{sec:maser}
No \water\ or \meth\ maser line emission was detected above $3\sigma$ in P6.
In P1, we detect two 22 GHz \water\ masers which we name as W1 and W2,
in decreasing order of brightness (white and black crosses in Fig. \ref{fig:p1.nh3}).
W1 is located at (RA, Dec)$_{\rm J2000}$ = (18:10:32.902, -19:22:23.339), close to the eastern border of the dark filament.
It has a flux of 3 Jy at \vlsr\,$=30.5$ \kms.
W2 is located at (RA, Dec)$_{\rm J2000}$ = (18:10:28.298, -19:22:30.759), coincident with P1-SMA1.
It has a flux of 0.25 Jy and is seen at \vlsr\,$=37.5$ \kms.
Both W1 and W2 show a typical maser spectrum with a single velocity component and has a narrow linewidth (FWHM $<$ 1 \kms).
The spectral profile is not resolved at the 0.3 \kms\ channel width.

In addition to the two water masers,
\cite{pillai06b} reported a 6.7 GHz class II \meth\ maser using the \atca\ (ATCA),
which we assign as M1. Located at
(RA, Dec)$_{\rm J2000}$ = (18:10:28.248, -19:22:30.45),
M1 is $0''.7$ (2500 AU) west of W2 (Figs. \ref{fig:p1.outflow},\ref{fig:p1.2mass},\ref{fig:p1.nh3_22}).
Unlike a single velocity component seen in the water masers,
M1 has multiple velocity components ranging from 22 to 34 \kms.
\cite{pillai06b} noted that M1 consists of six maser spots which are spatially distributed along a $0''.3$ north-south arc, and exhibits an ordered velocity field red-shifting from north to south. Based on the position and velocity distribution, \cite{pillai06b} suggested that the \meth\ maser spots trace a rotating Keplerian disc seen edge-on. Our detection of an East-West molecular outflow centred on W2 strongly supports this speculation (see \S~\ref{sec:outflow.p1}).

\begin{figure*}
\includegraphics[width=0.7\textwidth,angle=-90]{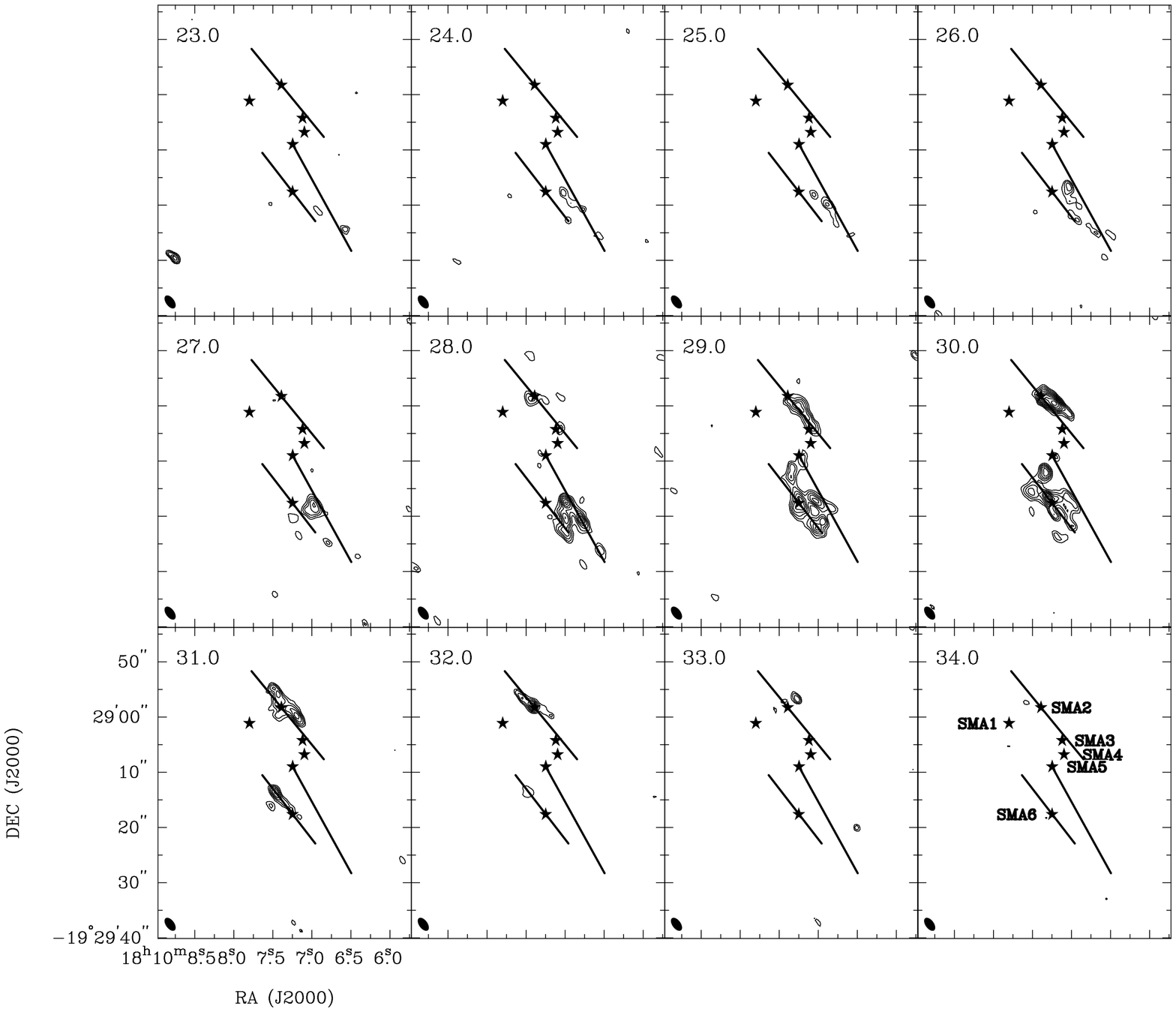}
\caption{
Channel maps of $\rm{CH_3OH}\,(4-3)$ in \gbpb.
{All panels share the same coordinates as labelled on the lower-left panel.}
Central velocity (\kms) of each channel is labelled on the upper left corner.
The contours are $\pm(3,4,5,6,...)\sigma$, where $\sigma = 25$\mjy\,\kms.
The stars denote six cores identified in \gbpb\ (SMA1--6, labelled in the last panel).
SMA2, SMA5, and SMA6 drive either bipolar or uni-polar outflows, as outlined by schematic lines.
The {filled} ellipse represents synthesized beam.}
\label{fig:p6.outflow}
\end{figure*}

\begin{table*}
\begin{minipage}{160mm}
\caption{Outflow Parameters \label{tab:outflow}}
\begin{tabular}{l cc cc cc cc cc}
\hline
{Parameter $^a$}
& \multicolumn{2}{c}{P1-SMA1}
& \multicolumn{2}{c}{P6-SMA2}
& {P6-SMA5}
& \multicolumn{2}{c}{P6-SMA6}
\\
& Blue & Red
& Blue & Red
& Blue
& Blue & Red
\\
\hline
Tracer							            	&SiO		&SiO		&\meth		&\meth		&\meth		&\meth		&\meth	\\
Fractional abundance [$X/{\rm H}_2$] $^b$
&$5\times10^{-10}$&$5\times10^{-10}$ 
&$5\times10^{-8}$ &$5\times10^{-8}$ &$5\times10^{-8}$ &$5\times10^{-8}$ &$5\times10^{-8}$\\
Excitation temperature \tex\ (K)            	&25   		&25   		&21    		&21    		&21    		&21    		&21       \\
Inclination angle $\theta$ (degree)$^c$ 	&77			&77 		&57.3    	&57.3    	&57.3    	&57.3    	&57.3       \\
Velocity range (\kms)     			    	&[18,28]	&[32,39]	&[22,29] 	&[31,35] 	&[22,29] 	&[22,29] 	&[31,35]   \\
Total mass $M$ (\msun)             	      		&2.0		&1.1		&1.1    	&0.4   		&2.1    	&1.2    	&0.1       \\
Momentum $P$ (\msun\,\kms)                     	&10.8/48.2	&4.6/20.7   &1.8/3.3    &0.8/1.5    &6.1/11.3   &4.1/7.5    &0.3/0.5     \\
Energy   $E$ (\msun\,km$^2$ s$^{-2}$)          	&40.0/791.0	&12.4/245.6 &1.8/6.2    &1.0/3.4    &12.3/42.3  &8.8/30.3   &0.4/1.4    \\
Lobe length $L_{\rm flow}$ (pc) 				&0.30/0.31	&0.30/0.31  &0.17/0.20  &0.07/0.08  &0.24/0.29  &0.28/0.34  &0.08/0.10     \\
Dynamical age $t_{\rm dyn}$ ($10^3$\,yr)          &24.9/5.7	&31.9/7.4   &20.8/13.3  &12.5/8.0   &30.2/19.4  &35.7/22.9  &15.8/10.1     \\
Outflow rate \Mout ($10^{-5}$\,\msun\,yr$^{-1}$) &8.0/34.8	&3.3/14.3   &5.4/8.4	&3.2/5.1    &6.9/10.8   &3.5/5.4    &0.7/1.1     \\
\hline
\end{tabular}

\medskip
\textbf{Note}:
\\$^a${Parameters corrected for inclination follow after the ``/''.}
\\$^b${Adopted abundances are based on \cite{Sanhueza2012_IRDC_chemi} and \cite{Leurini2007} for SiO and \meth, respectively.}
\\$^c${Angle between outflow axis and the line of sight, see \S\,\ref{sec:outflow.p1} and \S\,\ref{sec:outflow.p6}.}
\end{minipage}
\end{table*}

\subsection{Protostellar Outflows} \label{sec:outflow}

\subsubsection{Outflow driven by P1-SMA1} \label{sec:outflow.p1}
Our SMA observations clearly reveal a bipolar outflow oriented east-west in P1. The outflow is seen in many molecular tracers including CO, SO, SiO, H$_2$CO, and \meth,
but only in SiO do both the blue and red lobes appear; other tracers only reveal the blue (eastern) lobe. Fig. \ref{fig:p1.outflow} plots the blue shifted SiO emission (18--28\kms, blue contours) to the east, the SiO emission close to the systemic velocity (29--31\kms, gray background), and
red shifted SiO emission (32--39\kms, red contours) to the west,
in comparison with the SMA 880\um\ continuum (black contours).
A bipolar SiO outflow is clearly defined by the blue/red lobes with respect to the dust continuum,
which is probably produced in shocks by an underlying jet.
We schematically draw the axis of the outflow on the plot, and measure a position angle of $94\pm12^{\circ}$. The geometric centre of the outflow is close to P1-SMA1, and we speculate that the outflow driving source is a protostar embedded in the dust condensation P1-SMA1, likely the \#9 protostar identified by \cite{Henning2010}. (See more discussion on the driving source later in \S\,\ref{sec:p1.disc}).
Previous studies have shown indirect evidence of an outflow associated with P1, for example broadening of line wings, possible extended 4.5\um\ emission, and enrichment of outflow tracers \citep{carey2000,pillai06b,Leurini2007,Cyganowski2008,gomez2011}.
Our high-sensitivity, high-resolution, broad-band SMA observations directly reveal this outflow
in multiple tracers for the first time, providing critical support for a previously speculated outflow-disc system in P1 (\S\,\ref{sec:p1.disc}).

The SiO outflow extends 0.3 pc away from the protostar in the east-west direction (Fig. \ref{fig:p1.outflow}). While the blue lobe further propagates towards the eastern edge of the  main filament, it induced the \nh3 {emission peaks A, B and D,} and probably excited the water maser W1 (see first paragraph in \S\,\ref{sec:nh3.shock}). 
The projected separation of the \nh3 peaks are $l_{\rm AB} = 0.3$ pc, and $l_{\rm AD} = 1$ pc {(Fig.\,\ref{fig:p1.nh3})}, i.e.,
the molecular outflow extends at least 1 pc away from the driving source in the eastern lobe. 
Further to the east from D, there is no dense gas to be heated and shocked.
In the western lobe, however, there is no dense gas beyond the red shifted SiO lobe which is 0.3 pc from the protostar. Because the powering source is located near the edge of the dense filament, there is more dense gas in the eastern lobe to be heated (and therefore being detected) than in the western lobe. The special location and orientation of this outflow provides a unique case to study environment dependence of outflow chemistry, which deserves further study.

We compute outflow parameters assuming LTE and optically thin SiO emission in the line wings, following the formulas given in \cite{me11}. We adopt an abundance of [SiO/H$_2$] = $5\times10^{-10}$ based on recent chemistry surveys towards IRDCs by \cite{Sanhueza2012_IRDC_chemi}. The inclination angle of this outflow can be inferred based on the \meth\ masers discovered by \cite{pillai06b}. The masers outline (part of) an ellipse with an eccentricity of $\sim 0.2$. Suppose the masers trace a circular disc, we infer an inclination angle of $\sim 77^{\circ}$ between the axis of the disc (also the outflow jet) and the line of sight. Therefore, the disc is almost edge-on to us and the outflow jet is almost parallel to the plane of sky. Table \ref{tab:outflow} shows the derived outflow parameters with and without correction for inclination. The P1-SMA1 outflow is a massive outflow judging from its energetics, in comparison with other outflows emanating from high mass protostellar objects \citep{Beuther02c,zhang2005}.

\subsubsection{Outflows Driven by P6-SMA2,5,6} \label{sec:outflow.p6}
We also detect molecular outflows in P6. These outflows are seen in many molecular tracers including CO, SO, SiO, H$_2$CO, but are best seen in \meth. Fig. \ref{fig:p6.outflow} shows channel maps of \meth\ (4--3), where one can see that SMA2,5,6 drive relatively collimated outflows. Outflows associated with SMA2 and SMA6 show both blue and red shifted lobes extending about 0.1--0.3 pc, whereas the SMA5 outflow only shows a blue lobe approximately 0.3 pc long. All these outflows are oriented in a NE-SW direction, coincident with emission from the \nh3 \33 and higher transitions (Fig. \ref{fig:p6.nh3}). The outflow parameters (Table \ref{tab:outflow}) are also calculated in a similar way as for P1. The inclination angles for the P6 outflows are unknown. We list outflow parameters without correction and with correction for an inclination of $57.3^{\circ}$, the most probable value for a random distribution of outflow orientations \citep{Bontemps1996_inc,Semel2009_inc}. Compared with outflow P1-SMA1, the P6 outflows are slightly less energetic but are still comparable with other high-mass outflows \citep{Beuther02c,zhang2005}.

The cores in \gbpa\ and \gbpb\ exhibit similar characteristics (mass, size, and global mass reservoir) to those in \gapa. The cores driving powerful outflows are undergoing accretion to build up stellar mass. With a typical star formation efficiency of 30\% in dense gas and a standard stellar initial mass function, the $\sim 10^3$ \msun\ clumps will eventually form massive stars in some of the cores once the protostellar accretion is completed. At that time, these clumps will become massive star clusters. One key difference is that we detect copious molecular emission in the cores in \gbpa\ and \gbpb, which we can use to assess their evolutionary state ({\S~\ref{sec:chem}}).

\begin{table*}
\begin{minipage}{140mm}
\caption{Observed Molecular Lines $^a$ \label{tab:lines}}
\begin{tabular}{lllll}
\hline
{Rest Freq.$^b$}    & {Molecule}   & {Transition}    & {Clump $^c$} & {Remark}\\
{(GHz)}
\\
\hline
217.10498      & $\rm{SiO}$       & $5-4$ 			& P1, P6 & Outflow, Fig.\ref{fig:p1.outflow} \\
217.23853      & $\rm{DCN}$       & $3-2$			& P1 \\
217.82215      & $c-\rm{HCCCH}$   & $6_{1,6}-5_{0,5}$	& P1 \\ 
218.22219      & $\rm{H_2CO}$     & $3_{0,3}-2_{0,2}$	& P1, P6 & Outflow \\
218.32472      & $\rm{HC_3N}$     & $24-23$     	& P1 \\ 
218.44005      & $\rm{CH_3OH}$    & $4_{2,2}-3_{1,2}$	& P1, P6 & Outflow, Fig.\ref{fig:p6.outflow} \\
218.47563      & $\rm{H_2CO}$     & $3_{2,2}-2_{2,1}$	& P1, P6 & Outflow \\
218.90336      & $\rm{OCS}$       & $18-17$			& P1 \\
219.56037      & $\rm{C^{18}O}$   & $2-1$				& P1, P6 & Outflow \\
219.79828      & $\rm{HNCO}$      & $10_{0,10}-9_{0,9}$	& P1 \\
219.94943      & $\rm{SO}$        & $6_5-5_4$		& P1, P6 & Outflow \\
220.39868      & $^{13}\rm{CO}$   & $2-1$				& P1, P6 & Outflow \\
228.91047      & $\rm{DNC}$       & $3-2$				& P1 \\
229.75881      & $\rm{CH_3OH}$    & $8_{-1,8}-7_{0,7}$ 	& P1, P6 \\
230.53797      & $\rm{CO}$        & $2-1$				& P1, P6 & Outflow \\
231.22100      & $\rm{^{13}CS}$   & $5-4$				& P1 \\
\hline
337.06110      & $\rm{C^{17}O}$   & $3-2$				& P1, P6 \\
335.56021      & $\rm{^{13}CH_3OH}$   & $12_{1,11}-12_{0,12}$	& P1 \\ 
345.79599      & $\rm{CO}$        & $3-2$				& P1, P6 \\
346.52940      & $\rm{CH_3CHO}$   & $18_{17,1}-17_{17,0}$	& P1 \\ 
346.97089      & $\rm{CH_3CH_2CN}$   & $17_{8,9}-17_{7,10}$	& P1, P6 \\ 
346.99834      & $\rm{HCO^+}$     & $4-3$				& P1, P6 \\
346.99991      & $\rm{CH_3CHO}$   & $18_{7,11}-17_{7,10}$	& P1, P6 \\ 
\hline
\end{tabular}

\medskip
\textbf{Note}:
\\$^a${\, Lines observed above $3\sigma$ at bands 230\ghz\ and 345\ghz.}
\\$^b${\, Rest frequency obtained from Splatalogue (\url{http://splatalogue.net}).}
\\$^c${\, Clump in which the line is observed.}
\end{minipage}
\end{table*}

\begin{figure*}
\centering
\includegraphics[width=0.85\textwidth,angle=0]{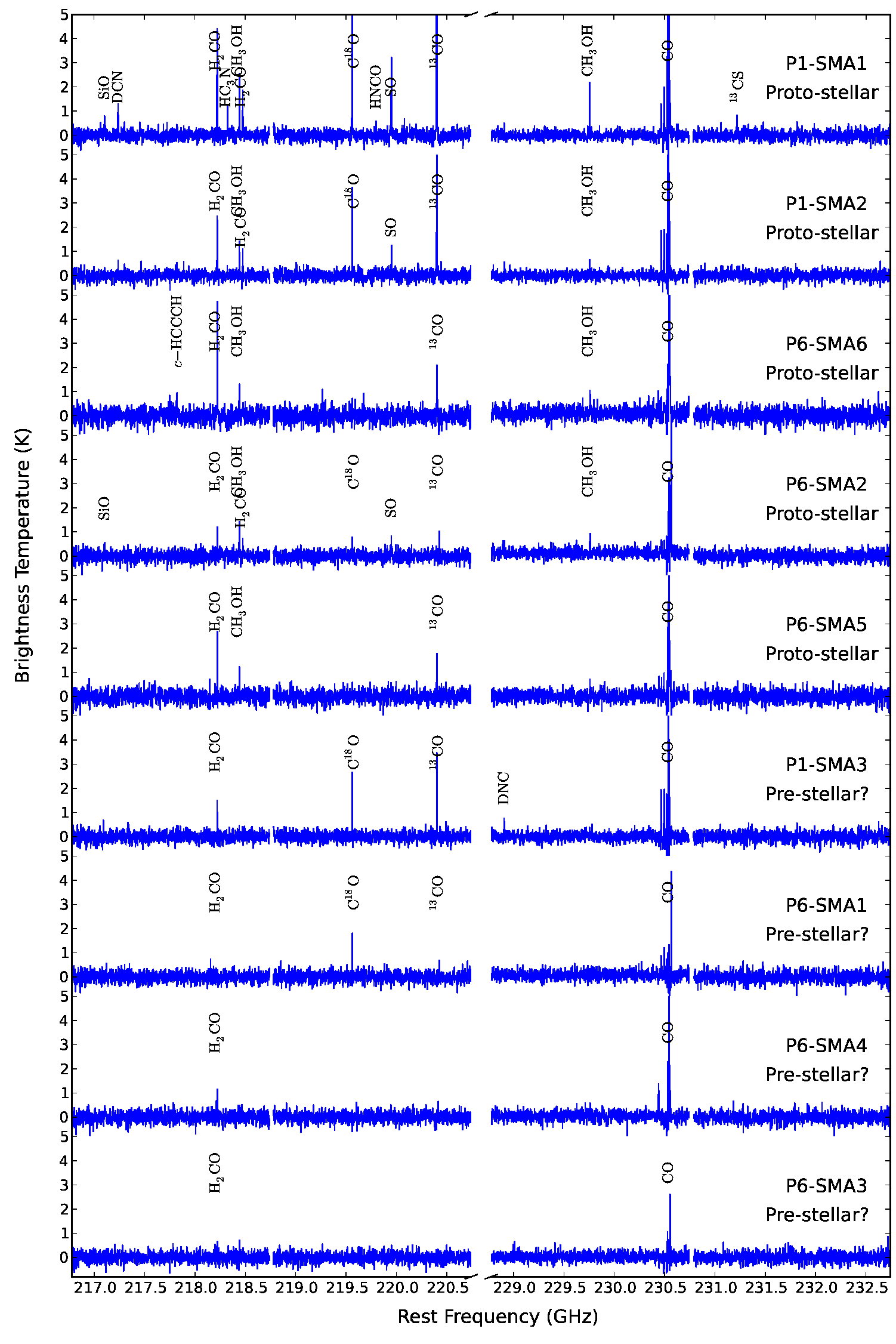}
\caption{
SMA 230\ghz\ spectra extracted from the ``major'' cores, plotted in inverse order of evolution. The brightness temperature has been corrected for the primary beam attenuation. The small gaps around 218.75 and 230.75 \ghz\ are due to our correlator setup, while the larger gap separates LSB and USB. Detected molecules are labelled.}
\label{fig:allspec}
\end{figure*}

\begin{figure*}
\centering
\includegraphics[width=.75\textwidth,angle=0]{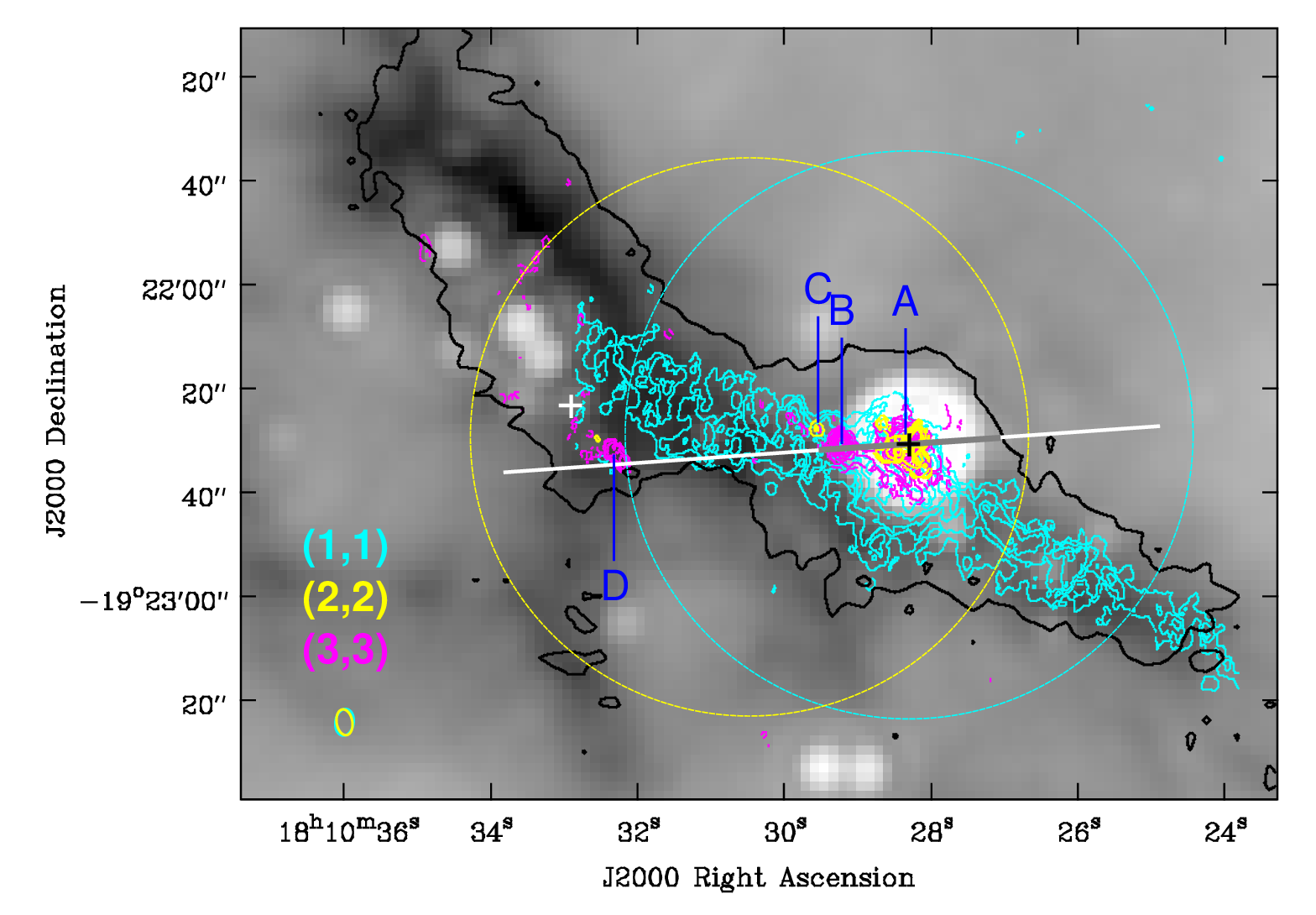}
\includegraphics[width=.21\textwidth,angle=0]{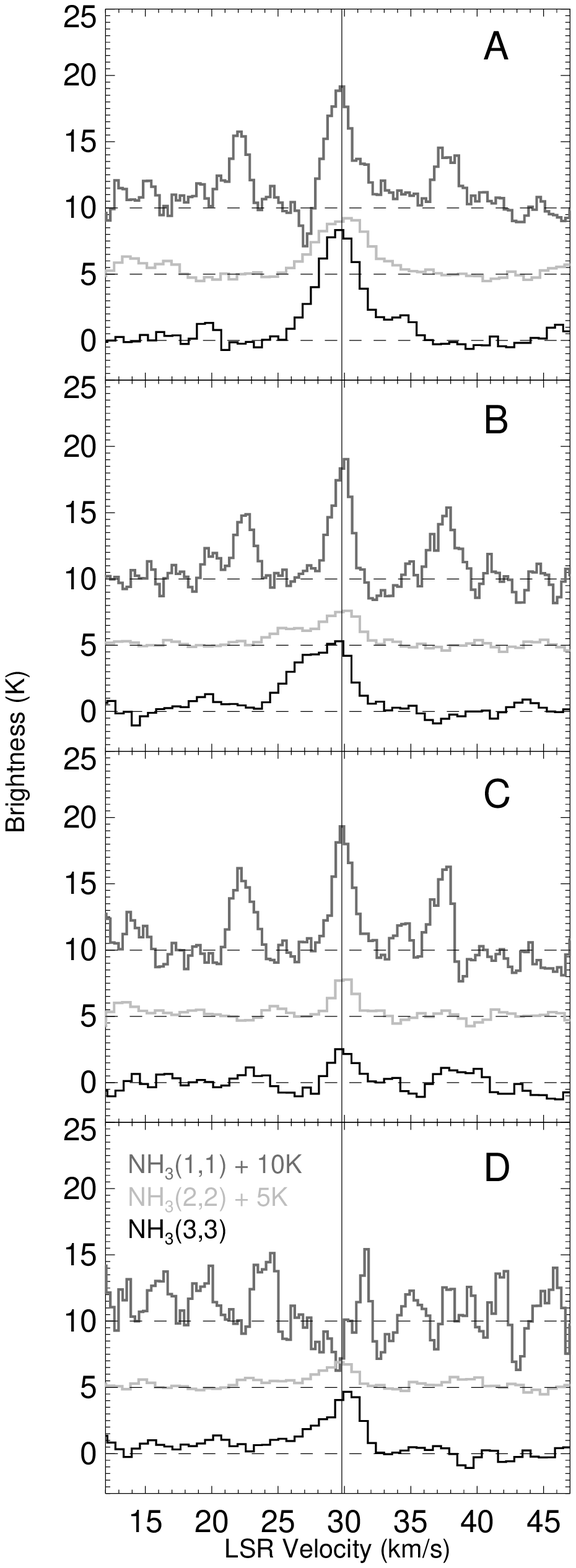}
\caption{
\textbf{Left:}
VLA observations of \gbpa: \nh3 \11, \22, \33
line emission integrated over the main hyperfine shown in
cyan, yellow, and pink contours, respectively.
The background is \spt\ 24\um\ image
{stretched from 22 (black) to 100 (white) MJy sr$^{-1}$}, and the black thick contour outlines
JCMT 850\um\ continuum emission at 0.31 Jy per $14''.5$ beam.
The contours for the \nh3 \11 emission are $(4,7,10,...,25)\sigma$, where $\sigma = 7.4$\mjy\,\kms;
The contours for the \nh3 \22 emission are $(4,6,8,...,24)\sigma$, where $\sigma = 2.8$\mjy\,\kms;
The contours for the \nh3 \33 emission are $(4,7,10,...,28)\sigma$, where $\sigma = 5$\mjy\,\kms.
The two (black and white) crosses denote the two water masers detected by VLA.
Dashed cyan/yellow circles represent two phase centres P1.I/P1.II (Tab. \ref{tab:obs})
and their relative FWHM primary beams.
Synthesis beams of the \nh3 images are shown as ellipses in the bottom left corner with the same color as the images. Four representative \nh3 emission peaks are labelled as A, B, C, and D.
The grey line is the same as in Fig. \ref{fig:p1.outflow} and represents the SiO (5--4) outflow driven by P1-SMA1. The white line is the extension of the grey line. {For scale, $l_{\rm AB} = 0.3$ pc, and $l_{\rm AD} = 1$ pc.}
\textbf{Right:}
\nh3 spectra extracted from emission peaks A--D as shown in the left panel.
The spectra are smoothed by a 2-point boxcar function to enhance features.
The brightness temperature has been corrected for the primary beam attenuation.
The vertical line represents the systematic velocity \vlsr $= 29.8$\kms.
}
\label{fig:p1.nh3}
\end{figure*}

\begin{figure*}
\centering
\includegraphics[width=.49\textwidth,angle=0]{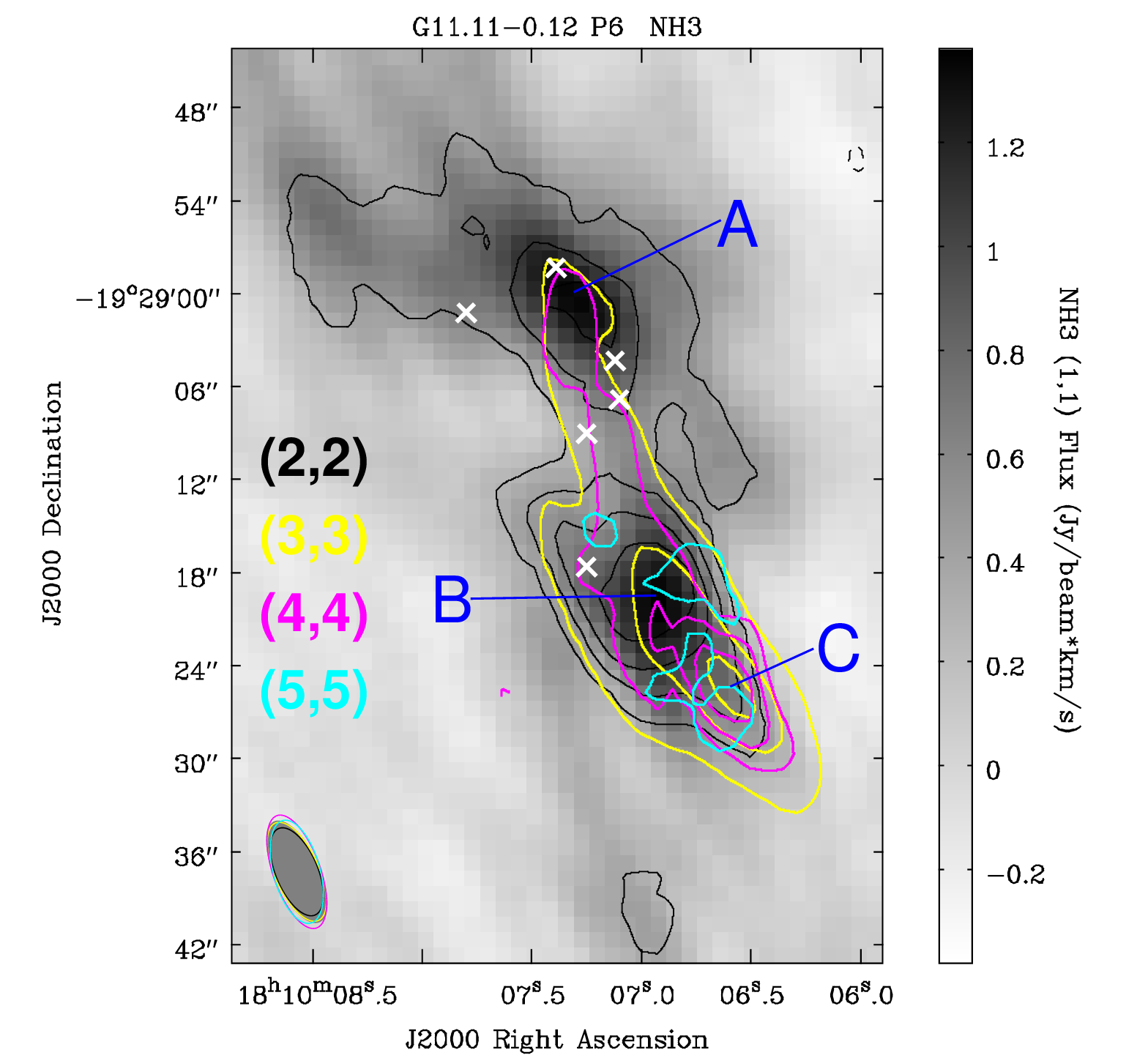}
\includegraphics[width=.49\textwidth,angle=0]{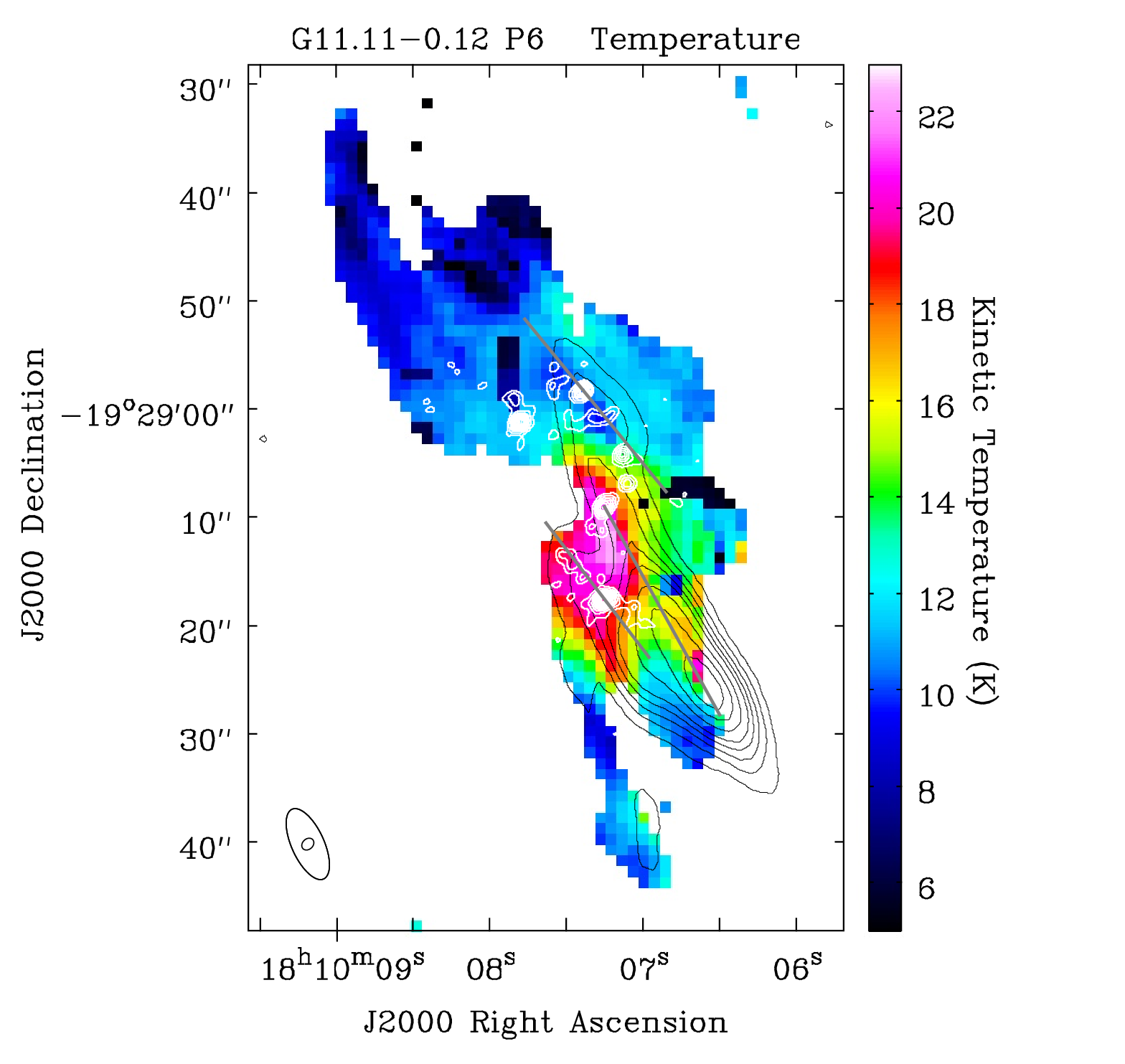}
\caption{
\textbf{Left:}
VLA observations of \gbpb: \nh3 \11, \22, \33, \44, and \55 integrated line emission plotted in gray scale, black, yellow, pink and cyan contours, respectively.
The \22 contours are $\pm (3,6,9,12)\sigma$, where $\sigma = 12$\mjy\,\kms;
the \33 contours are $\pm (6,20,30)\sigma$, where $\sigma = 25$\mjy\,\kms;
the \44 and \55 contours are $\pm (4,6,8,...)\sigma$, where $\sigma = 6$ and 8\mjy\,\kms, respectively.
The beams shown in the bottom left corner are color coded in the same way as the contours.
The white crosses mark the six dust cores SMA1--6. Selected regions (A,B,C) for OPR analysis are labelled.
\textbf{Right:}
Kinetic temperature map of \gbpb\ computed from \nh3 \11, \22, and \44 transitions. Overlaid are the SMA 880\um\ contours as in Fig. \ref{fig:con}(f), and the integrated \nh3 \33 emission with primary beam corrected shown in black contours
[$\pm (3, 6, 9, 12, 15, 20, 25, 30)\sigma$, where $\sigma = 25$\mjy\,\kms].
The small and large ellipses in the bottom left corner denote the resolution of the SMA image and the temperature map, respectively.
Grey lines denote the outflows as in Fig \ref{fig:p6.outflow}. The temperature map has been masked by 0.25 Jy ($3\sigma$) integrated \nh3 \11 emission.
}
\label{fig:p6.nh3}
\end{figure*}

\subsection{Chemical Differentiation} \label{sec:chem}

Our SMA broad-band observations covered a total of 16\ghz\ in two observing bands, and detected many molecular lines in P1 and P6. Table \ref{tab:lines} lists all detected lines and Fig. \ref{fig:allspec} plots spectra of the nine ``major'' cores. Besides CO isotopologues, P1 and P6 show several complex molecular lines like OCS, HNCO, $\rm{CH_3CN}$, and \meth, and molecules enriched by outflow shocks like SO and SiO. The number of detected lines vary from core to core and may reflect chemical evolution, despite that the cores are fragmented from the same parent clumps. Typical hot core lines $\rm{CH_3CN}$ and/or \meth\ are seen towards cores P1-SMA1,2, P6-SMA6,2,5, suggesting their protostellar nature. {This is consistent with the detection of protostellar outflows emanating from most of these cores.}
Deep \her\ 70\um\ image revealed point sources coincident with P1-SMA1 and P6-SMA6 and relatively diffuse 70\um\ emission in good agreement with our SMA 1.3\,mm images of P1 and P6, suggesting that P1-SMA1 and P6-SMA6 have developed an increased luminosity than their fellow cores \citep{Ragan2012_EPoS}.

The diagnosis in outflow, hot core lines, and 70\um\ emission show that P1-SMA1,2, P6-SMA6,2,5 are protostellar cores, while the other cores P1-SMA3, P6-SMA1,4,3 are likely of prestellar nature. Moreover, the line richness and strength reveal detailed chemical differentiation, likely reflecting an evolutionary sequence from core to core, as we plot in order in Fig. \ref{fig:allspec}. 
Ideally, one would compare cores with the same/similar mass, as a lower mass core can be more evolved but does not have detectable line emission. However, this does not seem to affect our results, since the most massive core P6-SMA1 is not most evolved, and the most evolved core P1-SMA1 is not most massive.
P1-SMA1 is in a ``warm core'' phase because it has not yet reached the hot core phase defined by \cite{Cesaroni2005_HMC}: temperature $\geq 100$ K, size $< 0.1$ pc, mass $10-1000$\msun, and luminosity $>10^4$ \lsun. The prestellar cores all show $\rm{H_2CO}$, a mid-product along the sequential hydrogenation from CO to \meth\ \citep{Zernickel2012}, thus they are slightly more evolved than the cores in IRDC clumps \gapa\ and \gcca\ where only $\rm{^{12}CO}$ is detected \citep{qz09,qz11,me11}. Distance effect cannot explain the difference, see discussion in \S~\ref{sec:evo.clumps}.
Within each core, the grouped condensations show similar chemical difference.

\subsection{\nh3 Emission and Temperature} \label{sec:nh3}
\subsubsection{Shock Heated \nh3} \label{sec:nh3.shock}

We detect ammonia emission in all the observed transitions both in P1 and P6.
Figure \ref{fig:p1.nh3} (left panel) shows the \nh3\ integrated images of P1 superposed on \spt\ 24\um\ and JCMT 850\um\ images.
The sensitive \spt\ 24\um\ image reveals details in the central part of the Snake Nebula:
a filamentary system that consists of a main NE-SW oriented dark filament and two minor filaments joining from the South. This configuration resembles the filamentary system discussed by \cite{Myers2009a} and may arise from compression of an elongated clump embedded in a thin cloud sheet, as seen in IRDC G14.225{\ndash}0.506 \citep{Busquet2013}. Dense gas traced by the JCMT dust image is mostly concentrated on the main filament,
and \nh3\ appears only on the main filament.
The \nh3\ \11 emission is relatively continuous, whereas \nh3 \22 and \33 emission are highly clumpy.
We identify four representative \nh3\ emission peaks A, B, C, and D, 
and plot the corresponding spectra in Fig. \ref{fig:p1.nh3} (right panel).
Peak A shows the strongest \nh3\ emission, and is associated with the IR point source (\# 9),
dust core P1-SMA1, and masers W2 and M1.
Following A, peaks B, C, and D are roughly aligned on a line eastward inside the main filament,
with D located near the eastern edge of the dense filament close to the water maser W1.
All peaks except C show broad line wings in all three transitions.
Peak A shows nearly symmetric blue and red line wings, whereas peaks B and D show
only the blue wings extending greater than 15 \kms\ from the systematic velocity.
Broad line wings, geometric alignment, and association with masers strongly suggest that peaks A, B, and D are associated with outflow activities.
{Indeed, these peaks are located on the extension of the blue lobe of the SiO outflow driven by P1-SMA1.
The line broadening increases with higher transition, and becomes more prominent in (2,2) and (3,3) than in (1,1). This suggests that a significant fraction of the higher excited (2,2) and (3,3) emission may arise from}
the passage of outflow shocks which releases the \nh3 molecules from dust mantle into the gas phase \citep{qz99,Nicholas2011_SNR}. The statistical equilibrium value for the fractional abundance ratio of ortho/para \nh3 is 1.0. But ortho-\nh3 ($K=3n$) is easier to release than para-\nh3 ($K \neq 3n$) because it requires less energy \citep{Umemoto1999}. Therefore, \nh3 \33 is expected to be enriched more than \22 and \11. The relative emission strength $I_{(3,3)} > I_{(2,2)} > I_{(1,1)}$ in D supports this scenario. We will test this quantitatively for P6 for which we have more transitions observed (\S\,\ref{sec:nh3.RD}).

Shock enhanced \nh3\ emission has been observed in a number of sources: a similar IRDC clump \gapa\ \citep{me12}, the high-mass disc/jet system IRAS 20126+4104 \citep{qz98,qz99}, the Orion BN/KL hot core \citep{Goddi2011_NH3}, and the low-mass L1157 outflow \citep{Umemoto1999}. We note that in a recent single dish study of a large EGO sample (shocked 4.5\um\ emission sources), \cite{Cyganowski2013_NH3} fit the \nh3 spectra with a fixed ortho-to-para ratio of unity. The fittings systematically underestimate the strength of \nh3 \33 (see Fig. 3 in their paper), which is suggestive to either (a) an elevated gas temperature traced by \nh3 \33, or (b) an enhanced ortho-\nh3, or a mixture of both. Both factors are the consequences of outflow activities.

Peak C is $3''$ offset from the outflow jet defined by A, B, and D. The spectral profiles are narrow and symmetric which indicates that the gas is unaffected by the outflow. This also suggests that the outflow is well collimated.

\nh3 emission also reveals an ordered velocity field towards P1. As an example, Fig. \ref{fig:p1.nh3_22} plots the moment 1 map of \nh3 \22 which clearly shows an NW-SE velocity field, varying more than 2\kms\ over 0.16 pc. We will discuss this velocity field in \S~\ref{sec:p1.disc}.
The velocity dispersion of \nh3 centres on P1-SMA1. On the main filament the overall velocity dispersion is about 0.8 \kms, which increases to 1.1 \kms\ towards P1-SMA1.

In P6, the \nh3 \11 and \22 emission follow the IR-dark dust ridge in general and are concentrated in two clumps: one lying between SMA2 and SMA3 and another southwest of SMA6, offset from the IR source (Fig. \ref{fig:p6.nh3} left panel). The \33, \44 and \55 emission lie on a slightly bent filament connecting the two \nh3 clumps and extending further towards southwest. While the \nh3 \11 and \22 are {less} affected by outflows, higher transitions are clearly associated with outflows (\S~\ref{sec:nh3.temp}). The velocity dispersion peaks on the two \nh3 clumps with values up to 0.9\kms. Towards the dust cores, the velocity dispersion varies from 0.4 to 0.7\kms\ with an average of 0.6\kms. These numbers are used in the fragmentation analysis (\S\,\ref{sec:frag}).

\subsubsection{\nh3 Temperature} \label{sec:nh3.temp}

{Metastable \nh3 inversion lines provide}
a robust thermometer for cold and dense gas in star formation regions \citep{Ho1983,Walmsley1983,Juvela2012_NH3}. To deduce the \nh3 temperature, we model the \11, \22, and \44 cubes simultaneously on a pixel-by-pixel basis, following the method developed by \cite{Rosolowsky2008_NH3}.
The model assumes LTE and Gaussian line profiles, and describes the spectra with five free parameters, including kinetic temperature, excitation temperature, \nh3 column density, velocity dispersion, and line central velocity.
The level population is governed by the rotational temperature of the \nh3 system, which is related to kinetic temperature with collision coefficients \citep{Danby1988}.
For details of the method see \cite{Rosolowsky2008_NH3}.
Only para species are included in the fitting to avoid any difference in the origin of ortho and para \nh3.
The three \nh3 images were restored in a common beam, and then corrected for primary beam attenuation before input for the fitting procedure.
Only pixels with $>3\sigma$ (0.25 Jy) in integrated \nh3 \11 emission are fitted; other pixels are masked out.

Figure \ref{fig:p6.nh3} (right panel) shows the fitted kinetic temperature map of P6. The temperature distribution shows a single relatively-high-temperature spot of $\gsim 20$ K located between SMA5 and SMA6, comparing to 10--15 K in other cores. This ``hot'' spot is also likely related with the outflows originated from SMA5 and SMA6 (\S~\ref{sec:outflow.p6}).
In P1, however, due to the outflow broadening, the fitting is inappropriate. 
Instead of fitting, we adopt the method used for \gapa\ \citep{me12}. 
This method estimates the rotational temperature by comparing the \nh3 \11 and \22 emission integrated over a 1.5\kms\ velocity range centred on the systematic velocity.
The rotational temperature approximates kinetic temperature very well in the regime of $\lsim 20$ K \citep{Ho1983,Walmsley1983}.

The resolutions of the \nh3 images and therefore the temperature map are high enough to resolve the cores but not the condensations. We assume all condensations in a given core share the same core-averaged temperature. A higher resolution map is needed to resolve the fine temperature structures associated with individual condensations. We list the temperature of each condensation in Table \ref{tab:cond}. Gas and dust are coupled in such a dense environment, so the gas temperature equals to the dust temperature, and is used to calculate the condensation masses. The error in the fitting is about 1 K across P6, while for P1 we estimate a 3 K error bar, similar to \cite{me12}.
The estimated temperature for the SMA cores and condensations range from 15 to 25 K in P1 and 10 to 21 K in P6. The upper bounds of the temperature range
are consistent with the \her\ SED estimates \citep{Henning2010,Ragan2012_EPoS}. This means that there is no evidence of significant external heating in this cloud. At the clump scale, we adopt an average temperature of 15 K for both P1 and P6 based on previous single dish studies \citep{pillai06b,Leurini2007}.

\section{Analysis and Discussion} \label{sec:discuss}
\subsection{Fragmentation Analysis} \label{sec:frag}

The hierarchical structures seen in Fig. \ref{fig:con} indicate fragmentation at different spatial scales: a clump with its initial physical conditions (density, temperature, turbulence, and magnetic fields) fragments into cores, and some cores continue to fragment into condensations with increased density and temperature. These structures resemble what we see in IRDC clump \gapa, thus we discuss the fragmentation of P1 and P6 as in \cite{me11}. First, we estimate the density and temperature averaged over clump and core scales for fragmentation analysis. 
On clump scale, Gaussian fitting to the JCMT 850\um\ image \citep{carey2000,Johnstone2003} yields an integrated flux of 8.7 Jy in a FWHM size of $40'' \times 25''$ ($0.7 \times 0.4$ pc) for P1. The average temperature of clump P1 is 15 K (\S~\ref{sec:nh3.temp}), leading to a clump mass of $1.2\times 10^3$\msun\ and a mean density of $8.6\times 10^4$\,cm$^{-3}$ for P1. For P6, the JCMT image is best fitted with a 2D elliptical Gaussian function with {7 Jy integrated flux in} FWHM $43'' \times 23''$ ($0.75 \times 0.4$ pc). Similarly, we obtain a clump mass of $9.3\times 10^2$\msun\ and a mean density of $7.6\times 10^4$\,cm$^{-3}$. On core scale, we fit the SMA 1.3\mm\ image shown in Fig. \ref{fig:con}(b,e) and compute core masses 10--92\msun\ and average core density $7.5\times 10^6$\,cm$^{-3}$. Calculated temperature and density on both scales, as well as the corresponding Jeans mass and length, are tabulated in Table \ref{tab:frag}. In this section we compare the observational results with theoretical predictions of Jeans fragmentation and cylindrical fragmentation, which are summarized in Fig. \ref{fig:frag}.

\begin{table*}
\begin{minipage}{115mm}
\caption{Hierarchical Fragmentation in {P1 and P6 $^a$}}
\label{tab:frag}
\begin{tabular}{lccc ccc}
\hline
  {Fragmentation Level}
& {$T$ $^b$}
& {$n$}
& {$M_{\rm Jeans}$}
& {$\lambda_{\rm Jeans}$}
& {$M_{\rm frag}$ $^c$}
& {$\lambda_{\rm frag}$ $^c$} 
\\
{}
& {(K)}
& {(cm$^{-3}$)}
& {(\msun)}
& {($10^{-2}$ pc)}
& {(\msun)}
& {($10^{-2}$ pc)}
\\
\hline
Clump$\rightarrow$cores
& 15  
& $8.0\times10^4$  
& 1.8
& 9.0  
& 10--92
& 7.5--14
\\
Core$\rightarrow$condensations 
& 10--25  
& $7.5\times10^6$  
& 0.1--0.4 
&0.8--1.2  
& 1.9--27.9 
& 1.6--8.3
\\
\hline
\end{tabular}

\medskip
\textbf{Note}:
\\$^a${{Listed are average or range of properties measured in all the clumps, cores, or condensations. The clump properties are measured from JCMT data, and others from SMA data.}}
\\$^b${Coupled dust and gas temperature estimated from multiple \nh3 lines, see \S~\ref{sec:nh3.temp}.}
\\$^c${Observed mass/separation of/between ``fragments'' (core or condensation).}
\end{minipage}
\end{table*}

\subsubsection{Jeans Fragmentation} \label{sec:str.jeans}

If the fragmentation is governed by Jeans instability, the initially homogeneous gas characterized by particle density $n$ (or equivalently, mass density $\rho$)
\footnote{Particle number density $n$ and mass density $\rho$ are related as $\rho = \mu m_\mathrm{H} n$, where $\mu = 2.37$ is the mean molecular weight per ``free particle'' (H$_2$ and He, the number of metal particles is negligible). We use particle number density throughout this paper. Note this is different to the H$_2$ number density which counts hydrogen molecules only and thus a higher molecular weight of 2.8 should be applied. See discussion in \cite{Kauffmann2008}.
For a molecular cloud made with $N(\mathrm{H})/N(\mathrm{He}) = 10$ and negligible metals, $n_{\mathrm{particle}}:n_{\mathrm{H_2}} = 1.2$.}
and temperature $T$ has a Jeans length of
\begin{equation}
\lambda_{\rm J} = c_s
\left( \frac{\pi}{G \rho} \right)^{1/2} =
0.066 ~ {\rm pc} \,
\left( \frac{T}{10\, \mathrm{K}} \right) \, 
\left( \frac{n}{10^5 ~ \rm cm^{-3}} \right)^{-1/2},
\label{eq:jeans-length}
\end{equation}
where $G$ is the gravitational constant, and $c_s = \sqrt{\frac{kT}{\mu m_\mathrm{H}}}$ is the speed of sound at temperature $T$. The Jeans mass is the gas mass contained in a sphere with an radius of $\lambda_{\rm J}/2$:
\begin{equation}
M_{\rm J} = \frac{\pi^{5/2} c_s^3}{6 \sqrt{G^3 \rho}} =
0.877 \, M_{\odot} \,
\left( \frac{T}{10\, \mathrm{K}} \right)^{3/2} \, 
\left( \frac{n}{10^5 ~ \rm cm^{-3}} \right)^{-1/2} \, ,
\label{eq:jeans-mass}
\end{equation}

In the above two equations, the sound speed $c_s$ represents pressure from thermal motions. Massive clumps often contain non-thermal motions likely dominated by micro turbulence. If the internal pressure of the gas is dominated by turbulence, then $c_s$ is replaced by the turbulent linewidth which is well approximated by the velocity dispersion measured from dense gas tracers like \nh3. As Table \ref{tab:frag} shows, at both fragmentation levels (clump or core), the masses of the observed ``fragments'' (core or condensation, respectively) are consistent with the prediction of Jeans fragmentation only if we account for the turbulent pressure; otherwise, thermal pressure alone predicts much smaller fragment mass. This suggests that turbulence dominates over thermal pressure in the hierarchical fragmentation, the same as we found in IRDC clump \gapa\ \citep{me11}, and in contrast to low-mass star formation regions \citep{Lada2008_Pipe}.

\subsubsection{Cylindrical Fragmentation} \label{sec:str.cylinder}

Similar to the case of \gapa, the cores in \gbpb\ and perhaps also in \gbpa\ appear to be regularly spaced along a ``cylinder'', although these ``cylinders'' are not straight. The so-called ``sausage instability'' of a gas cylinder has been suggested to explain such fragmentation, as first proposed by \cite{Chandra1953} and followed up by many others \citep[e.g.,][see also discussion in \citealt{Jackson2010, me11}]{Ostriker1964_FL,Nagasawa1987_FL,Bastien1991_FL,Inutsuka1992_FL,Fischera2012_FL}.
In an isothermal gas cylinder, self gravity overcomes internal pressure (represented by $\sigma$) once the mass per unit length along the cylinder (or linear mass density) exceeds a critical value of
\begin{equation}
(M/l)_{\rm crit} = 2\sigma^2/G = 465\,\left(\frac{\sigma}{1~\rm km\,s^{-1}}\right)^2 M_\odot\,{\rm pc}^{-1}.
\label{eq:ml-cylinder}
\end{equation}
Under this condition, the cylinder becomes gravitationally unstable and fragments into a chain of equally spaced fragments with a typical spacing of
\begin{equation}
\lambda_\mathrm{cl} = 22v(4\pi G\rho_c)^{-1/2} = 1.24\, {\rm pc}
\left( \frac{\sigma}{1 ~ \rm km \, s^{-1}} \right)
\left( \frac{n_c}{10^5 ~ \rm cm^{-3}} \right)^{-1/2}.
\label{eq:l-cylinder}
\end{equation}
In the above two equations, $\rho_c$ or $n_c$ is the gas density at the centre of the cylinder.
The fragment mass is therefore
\begin{equation}
M_\mathrm{cl} = (M/l)_{\rm crit} \times \lambda = 575.3 M_\odot
\left( \frac{\sigma}{1 ~ \rm km \, s^{-1}} \right)^3
\left( \frac{n_c}{10^5 ~ \rm cm^{-3}} \right)^{-1/2}.
\label{eq:m-cylinder}
\end{equation}
Similarly, if the cylinder is supported by thermal pressure, $\sigma$ is the sound speed $c_{\rm s}$;
if, on the other hand, it is mainly supported by turbulent pressure, 
then $\sigma$ is replaced by the velocity dispersion.

Assuming the initial temperature of the cylinder that represents clumps P1 and P6 is the average clump temperature (15 K), then the corresponding sound speed is $c_s = 0.23$\kms, while the average velocity dispersion is $\sigma \sim 0.7$\kms. According to Eq. \ref{eq:ml-cylinder} these lead to a linear mass density of $298 ~ M_{\odot}\,{\rm pc}^{-1}$ for turbulence support which is in agreement with the observed core masses in P1 and P6, while for thermal support this value becomes 25 $M_{\odot}\,{\rm pc}^{-1}$, inconsistent with the observations.
On the other hand, the projected separations of the cores are 0.08--0.14 pc (Table \ref{tab:frag}). We estimate the central density of the cylinder on the verge of collapse to be the average of the clump density and core density, which amounts to $n=3.75\times 10^6 \rm{cm}^{-3}$. We calculate a separation of 0.1 pc for turbulent support and 2 pc for thermal support. Clearly the observations favor the former against the latter scenario.
{On larger scales, taking the entire Snake filament as a cylinder, the mass/separation between the clumps is also consistent with a cylindrical fragmentation. Adopting an average velocity dispersion of 0.7\kms\ and an initial central density of $1\times10^4$\cmc\ \citep{Kainulainen2013_Snake}, the predicted clump mass and separation are 624\msun\ and 2.7 pc, in agreement with the measured average values of 1100\msun\ and 3 pc.}

Similar turbulence dominated fragmentation and cylindrical collapse have been observed by SMA and VLA in another IRDC \gc\ \citep[][see for example Fig. 1 in their paper]{qz11}. There, in the clump \gccb, dust cores SMM2,3,4 lie aligned in a north-south filament also seen in \nh3. The filament may extend further north where an \water\ maser and an \nh3 core coincide with each other (unfortunately this is far beyond the coverage of the SMA observations), and further south to a 24\um\ point source. All these features together form a north-south, 1.5 pc filamentary structure. The core mass and separation support a fragmentation and cylindrical collapse similar to \gbpa, \gbpb, and \gapa.

\begin{figure*}
\centering
\includegraphics[width=0.98\textwidth,angle=0]{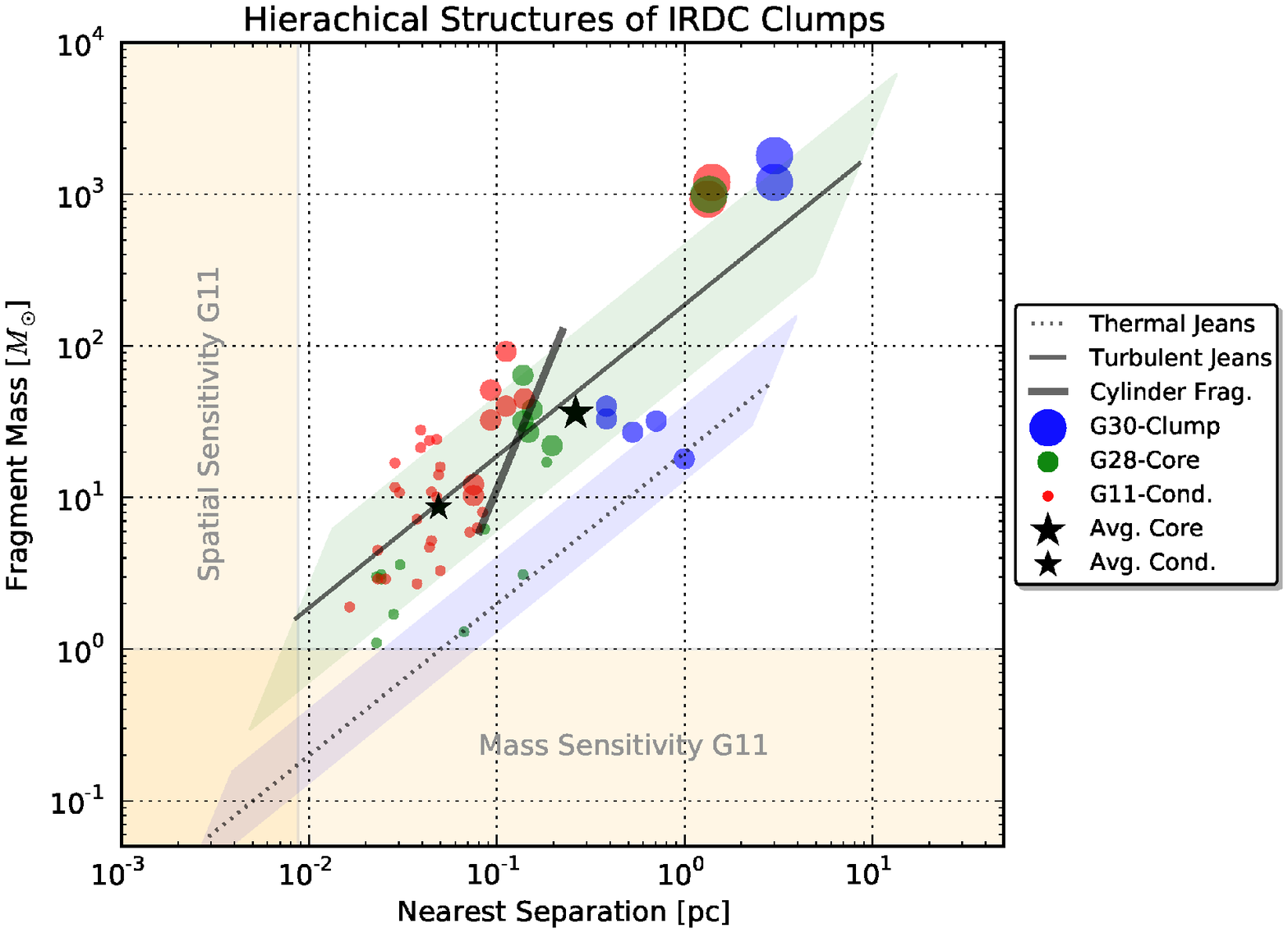}
%
\caption{
Fragment mass versus projected separation to the nearest fellow fragment. Circles filled with various sizes denote clump, core, and condensation. Sources are color coded as \gbpa\ and \gbpb\ (G11, red; {this work}), \gapa\ (G28, green; {\citealt{me11,me12}}), \gcca\ and \gccb\ (G30, blue; {\citealt{qz11}}). The orange shaded regions show the sensitivity and resolution limit of the G11 observations. The stars denote average observed properties of all the cores or condensations.
The dotted line shows thermal Jeans fragmentation with $T=15$ K and $n = [10^2, 10^8]$ \cm3, and the blue shaded region corresponds to the same density range but with $T = [10, 30]$ K.
The thin solid line shows turbulent Jeans fragmentation with $\sigma = 0.7$\kms\ and the same density range, and the green shaded region corresponds to the same density range but with $\sigma = [0.4, 1.1]$\kms.
The thick solid line shows cylinder fragmentation with central density $n_c = 3.75 \times 10^6$ \cm3\ and $\sigma = [0.4, 1.1]$\kms.
This figure shows clearly that the fragmentation from clump to cores and from core to condensations are dominated by turbulence over thermal pressure.
See text in \S\,\ref{sec:frag} for discussion.
}
\label{fig:frag}
\end{figure*}

The fragmentation analysis is summarized in Fig. \ref{fig:frag}. For a given core, we plot its mass against the separation to the nearest core. Similarly, clumps and condensations are also plotted. Fragments in different IRDCs are plotted with different colors: \gb\ in red, \ga\ in green, and \gc\ in blue.
The two clumps in \gc\ overlap completely along the line of sight, and are apparently isolated. We assign a nominal separation of 3 pc (the approximate long axis of the \gcca\ filament) for illustration only.
The shaded regions represent Jeans mass and length (Eqs. \ref{eq:jeans-length},\ref{eq:jeans-mass}) for a range of physical properties. For thermal Jeans fragmentation (shaded blue), the mass and separation are determined by temperature and density. The adopted temperature range $T=[10, 30]$ K is observed in these IRDCs, and the density $n = [10^2, 10^8]$ \cm3 is a wide range representative for molecular clouds to dense condensations. For turbulent Jeans fragmentation (shaded green), the temperature is replaced by the observed velocity dispersion range $\sigma = [0.4, 1.1]$\kms. Overlapped on the shaded regions are lines for a representative temperature (15 K) or an average velocity dispersion (0.7\kms). In addition, cylindrical fragmentation (Eqs. \ref{eq:l-cylinder},\ref{eq:m-cylinder}) for central density $n_c = 3.75 \times 10^6$ \cm3\ and $\sigma = [0.4, 1.1]$\kms\ is plotted as a thick solid line.
We find that most cores and condensations are located within the shaded green region corresponding to the turbulent fragmentation, and the average properties of cores and condensations (denoted by black filled stars) lie almost along the thin solid line. The cores are also well represented by the thick solid line corresponding to cylindrical fragmentation. On the other hand, thermal Jeans fragmentation (shaded blue) predicts much smaller fragment masses than observed.
This figure shows that the hierarchical fragmentation observed in these IRDC clumps are the same in nature, i.e., turbulence dominated fragmentation, and furthermore, the fragmentation from the clump to the core scale can be well explained by turbulence supported gravitational collapse of an isothermal cylinder. Our combined SMA and VLA observations have resolved similar fragmentation and cylindrical collapse in four IRDC clumps that are in different evolutionary stages (\S\,\ref{sec:evo.clumps}), suggesting that turbulence supported fragmentation is common in the early stages of massive star formation. There is a lack of direct measurements of magnetic fields in IRDC clumps. However, magnetic fields at a few mG levels are reported in protocluster regions such as G31.41 \citep{girart2009}, indicating the importance of the B field relative to turbulence.

\subsection{\nh3 Ortho/Para Ratio} \label{sec:nh3.RD}

\begin{figure*}
\includegraphics[width=0.33\textwidth,angle=0]{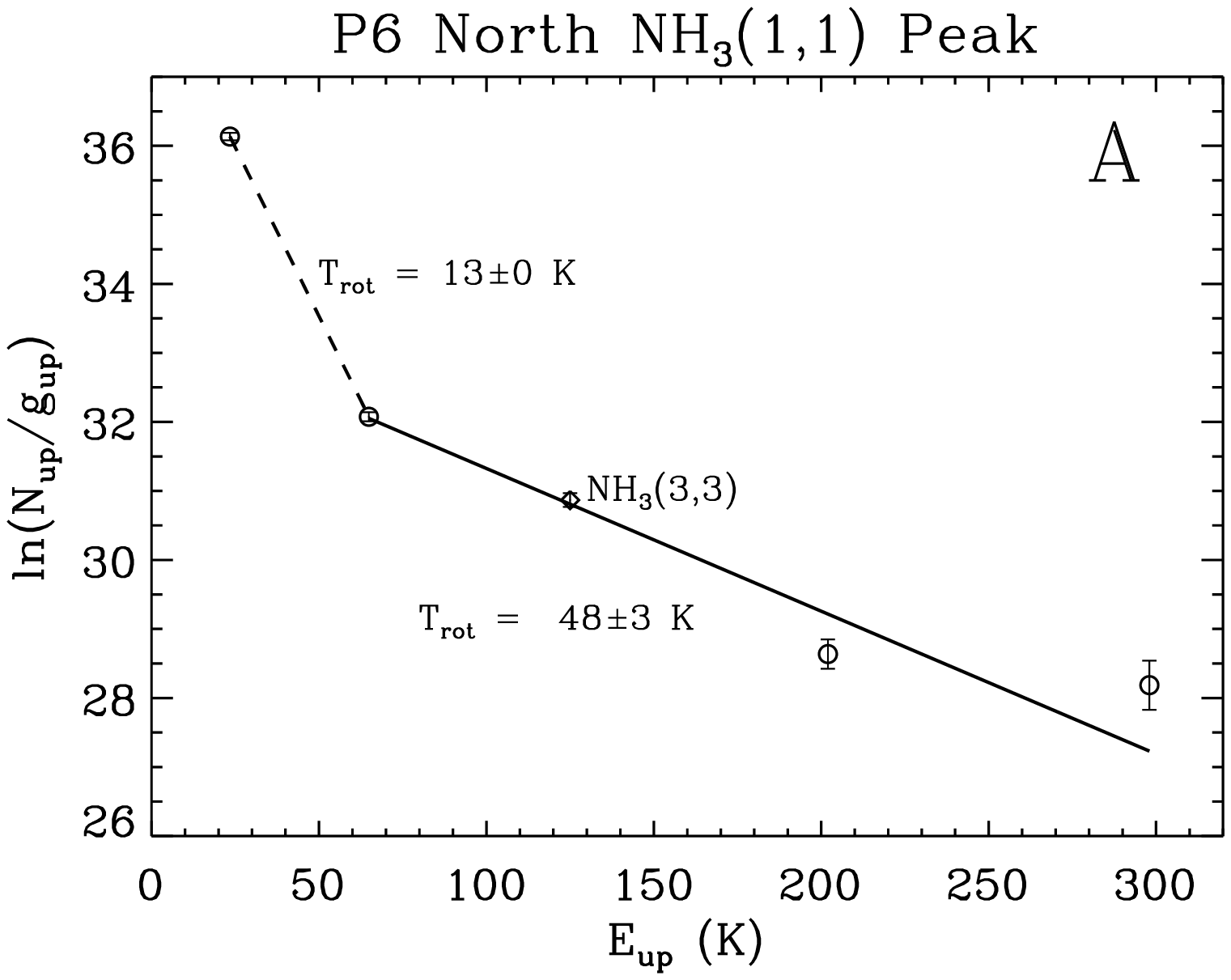}
\includegraphics[width=0.33\textwidth,angle=0]{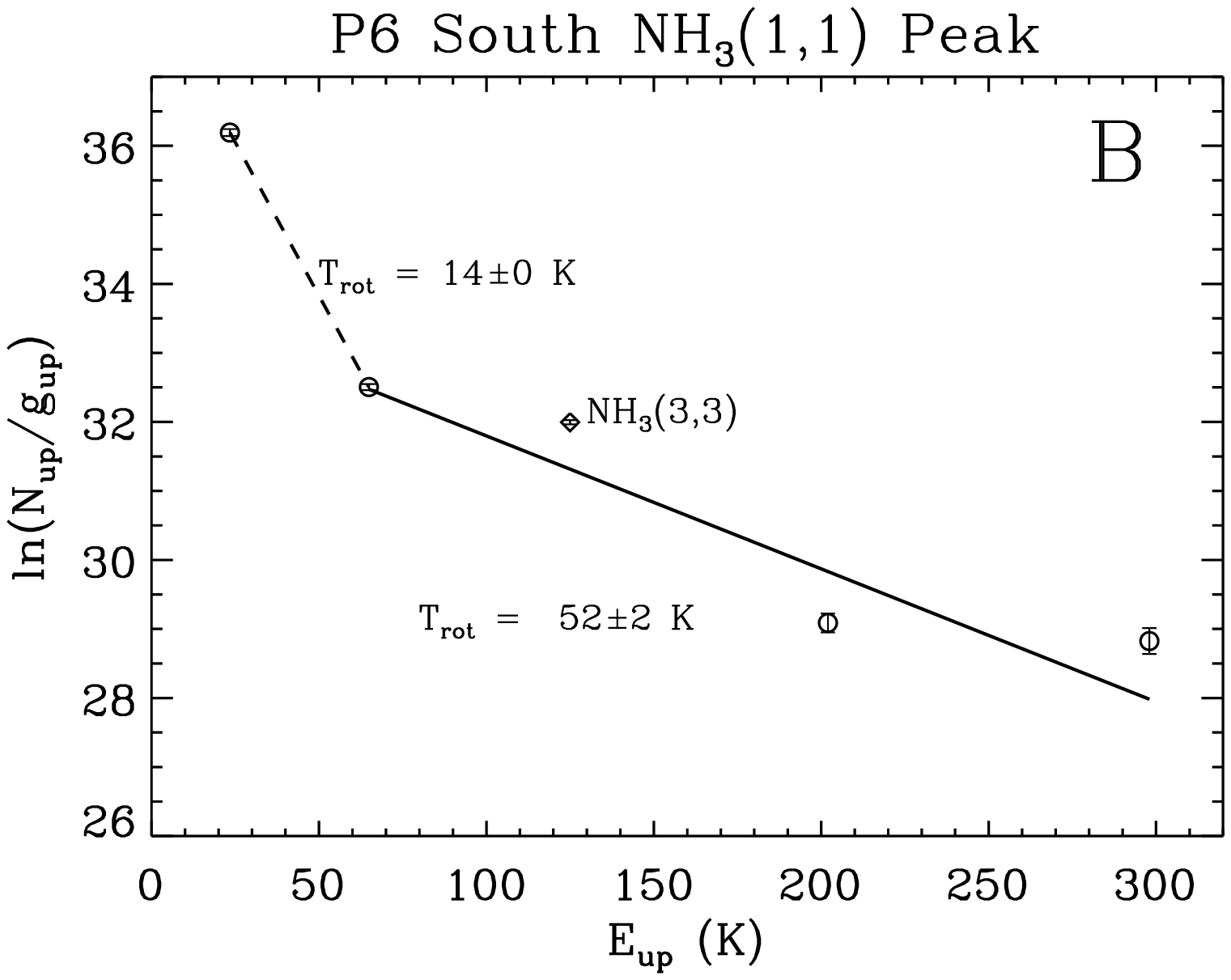}
\includegraphics[width=0.33\textwidth,angle=0]{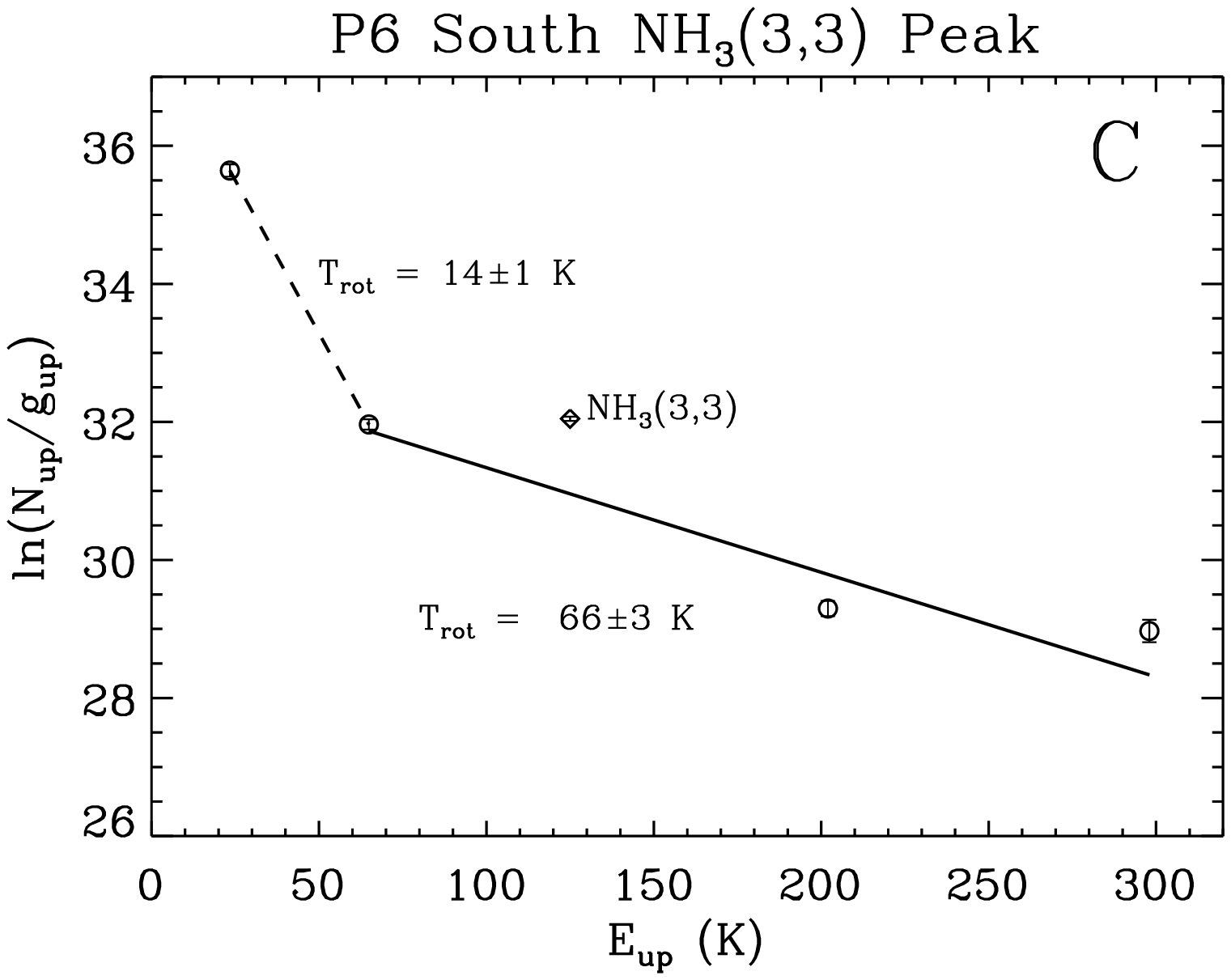}
\caption{\nh3 rotational diagrams of selected regions in P6 (see locations in Fig. \ref{fig:p6.nh3}). The error bars correspond to $1\sigma$ in the data points. The data are fitted with two temperature components, 
{one fits \22, \44, and \55 (solid lines) and another fits the residual of \11 and \22 after subtraction from the first fit. The dashed lines connect \11 and \22 emission summed from the two fits.}
The corresponding rotational temperatures are labelled for each fit. Other fitting results including the column density and OPR are reported in Table \ref{tab:opr}.
}
\label{fig:p6.o/p}
\end{figure*}

\begin{table*}
\centering
\begin{minipage}{140mm}
\caption{\nh3 Ortho/Para Ratio (OPR) {Associated with Outflows in P6} \label{tab:opr}}
\begin{tabular}{ccc cc}
\hline
{Region $^a$}
& {OPR}
& {$N_{\rm ortho}$}
& {$N_{\rm para}$}
\\
{}
& {}
& {($10^{16}$\cms)}
& {($10^{16}$\cms)}
\\
\hline
A
& $1.1 \pm 0.4$
& $1.3\pm0.6$
& $1.2\pm0.1$
\\
B
& $2.0 \pm 0.4$
& $3.6\pm0.9$
& $1.8\pm0.1$
\\
C
& $3.0 \pm 0.7$
& $3.3\pm1.1$
& $1.1\pm0.1$
\\
\hline
\end{tabular}

\medskip
\textbf{Note}:
\\$^a${ See locations in Fig. \ref{fig:p6.nh3} and parameter fitting in Fig. \ref{fig:p6.o/p}.}
\end{minipage}
\end{table*}

Quantitatively, the enrichment of ortho- over para-\nh3\ can be quested by comparing the fractional abundance ratio, [ortho/para], or OPR. We estimate the OPR by means of rotational diagram \citep{Turner1991_RD, Goldsmith1999_PD}. This technique requires measurement of many transitions so we only apply to P6.
The Boltzmann distribution characterized by an excitation temperature \trot\ leads to a linear equation
\begin{equation}
\begin{split}
\mathrm{ln}\left( \frac{N_\mathrm{u}}{g_\mathrm{u}} \right) 
&= 
\mathrm{ln}\left( \frac{N_\mathrm{tot}}{Q_\mathrm{rot}} \right) -
\frac{E_\mathrm{u}}{kT_\mathrm{ex}} \\
&= 
a+bE_\mathrm{u} \, ,
\end{split}
\end{equation}
where $N_\mathrm{u}$, $E_\mathrm{u}$, $g_\mathrm{u}$ are the column density, energy, and total degeneracy, of the upper state respectively, $N_\mathrm{tot}$ is the total column density summing up all transitions, 
$Q_\mathrm{rot}$ is the rotational partition function at \trot,
and $k$ is the Boltzmann constant.
In a Boltzmann plot, the slope $b$ is related to the rotational excitation temperature as
$T_\mathrm{rot} = -1/b$,
and the intercept $a$ is related to the total column density as
$N_\mathrm{tot} = Q_\mathrm{rot}e^a$.
The OPR is inferred by comparing the observed \nh3 \33 intensity to the fitted intensity should OPR equals unity:
$\mathrm{OPR} = {N^\mathrm{obs}_\mathrm{tot}}/{N^\mathrm{fit}_\mathrm{tot}} = \mathrm{exp}(a^\mathrm{obs} - a^\mathrm{fit})$.

As long as the Rayleigh-Jeans approximation ($h\nu \ll kT_\mathrm{rot}$) holds, and \trot $\gg$ \tbg, the column density can be computed from the observed intensity:
\begin{equation}
\begin{split}
N_\mathrm{u} 
&= 
\frac{3k}{8\pi^3 \nu S\mu^2} C_\tau \int T_\mathrm{B}\,dv \\
&=
\frac{7.75 \times \,\,10^{13} \mathrm{cm^{-2}}}{\nu \,\mathrm{(GHz)}} \,
\frac{J(J+1)}{K^2}
C_\tau
\int T_\mathrm{B}\,dv \,\mathrm{(K\,km\,s^{-1})}
\end{split}
\label{eq:Nu}
\end{equation}
where $\nu$ is the rest frequency of the transition, 
$\mu = 1.468$ Debye (1 Debye $=10^{-18}$ esu\,cm in cgs units) is the dipole moment for \nh3,
$S=\frac{K^2}{J(J+1)}$ is the line strength for $(J,K) \rightarrow (J,K)$ transitions,
{and $C_\tau = \tau/(1-e^{-\tau})$ is a correction factor for optical depth \citep{Goldsmith1999_PD}.
The optical depth $\tau$ is determined by comparing hyperfine line ratios \citep{Ho1983,LiD2003-Orion}. We find transitions \22\ and higher are optically thin.
}

We make rotational diagrams (Fig.~\ref{fig:p6.o/p}, Table \ref{tab:opr}) for three representative regions coincident with the outflows in P6 (\S\,\ref{sec:outflow.p6}): the northern \nh3 \11 peak (A), the southern \nh3 \11 peak (B), and the southern tip of the \nh3 \33 emission (C). These regions (labelled in Fig. \ref{fig:p6.nh3}) are chosen to (a) have a high signal-to-noise ratio and (b) represent different locations along outflows: region A overlays with the driving source of outflow SMA2, region B is close to SMA6, and region C is farther away but along the direction of outflows SMA5 and SMA6. Integrated \nh3 emission is extracted from a circular region with a diameter of $\sim 7''$, corresponding to the major axis of the largest beam of the \nh3 images.
{
Obviously, the diagrams (Fig.~\ref{fig:p6.o/p}) cannot be fitted well with a single temperature because \11 traces a lower temperature than other transitions. We fit the diagrams with two temperature components in the following approach: (i) fit one temperature to \22, \44, \55; (ii) subtract the fit from \22 and \11; and then (iii) fit another temperature to the residual of \11 and \22. This approach is chosen keeping in mind that emissions from \22 and higher transitions are dominated by outflow heating, judging from the broad ling wings.}
Indeed, the fit to \11 and \22 results {13--14 K} which likely traces the dense envelope, and a fit to \22, \44, \55 results a much higher temperature of {48--66 K} due to outflow heating. We assume that \nh3 \33 has the same temperature as the second fit, and find OPRs of $1.1\pm0.4$ (A), $2.0\pm0.4$ (B), and $3.0\pm0.7$ (C). The results indicate that (i) ortho-\nh3 is enhanced in all these regions, and (ii) the enhancement increases along the outflow downstream. This is consistent with the idea that \nh3 molecules have been released to gas phase from dust grains by outflow shocks, and that ortho-\nh3 is preferentially desorbed than para-\nh3. The \nh3 \33 spectra in P6 and P1 are free of maser lines, so the high OPRs are not due to masing emission. Ammonia OPRs greater than 1.0 (up to 1.6) have been reported in the Galactic centre \citep{Nagayama2009_OPR} and Galactic star formation regions \citep[L1157 and Orion][]{Umemoto1999,Goddi2011_NH3}. Even higher OPRs ($>6$) have been reported in starburst galaxies \citep{Takano2002_OPR}. OPR has also been found to be $<1.0$ \citep{Faure2013_OPR}.

\begin{figure*}
\centering
\includegraphics[width=0.6\textwidth,angle=0]{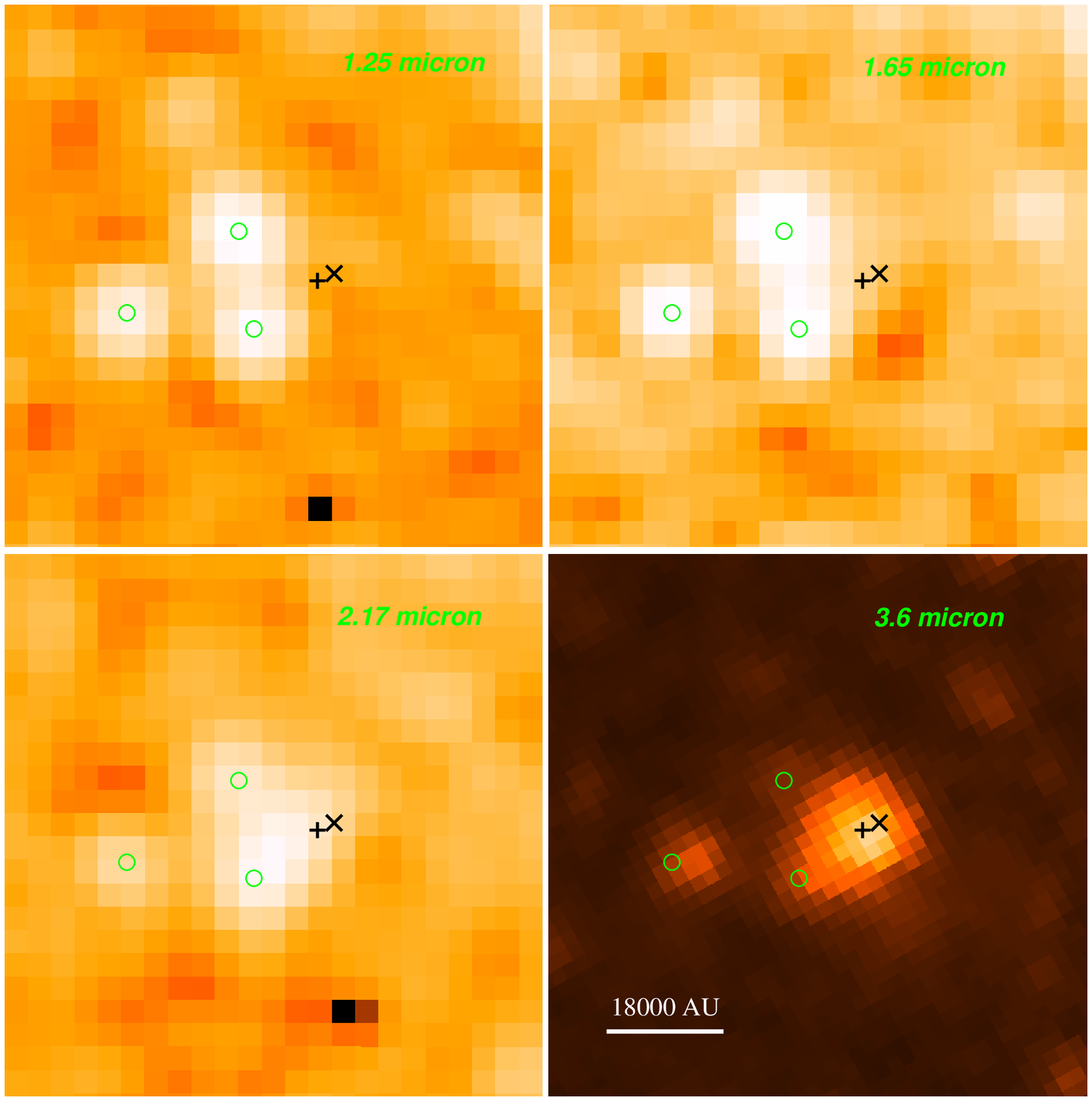}
\caption{
Infrared images of \gbpa: 2MASS  $J, H, K_s$ bands and \spt\ 3.6\um.
Labelled symbols denote \water\ maser W2 ($+$), class II \meth\ maser M1 ($\times$), and the three 2MASS point sources ($\circ$).}
\label{fig:p1.2mass}
\end{figure*}

It is noteworthy that our analysis has adopted a standard technique used in many similar studies \cite[e.g.,][]{Umemoto1999,Henkel2000_OPR,Mauersberger2003_OPR,Ao2011_OPR}, and presents comparable results. Ideally, a more sophisticated approach would require data from transitions up to \66 (i.e., including at least two data points from ortho-\nh3) and radiative transfer modelling incorporated with a temperature distribution.

\subsection{P1-SMA1: an Outflow/disc System in a Proto-binary?} \label{sec:p1.disc}

The east-west outflow corresponds surprisingly well with the north-south edge-on disc suggested by \cite{pillai06b} based on \meth\ masers. Without direct evidence of outflow, \citeauthor{pillai06b} hypothesized an outflow perpendicular to the disc which is responsible for illuminating the small dust knots seen as three point sources in 2MASS images (Fig. \ref{fig:p1.2mass}). \citeauthor{pillai06b} further speculated a geometric orientation of this hypothesized outflow-disc system, see Fig. 6 in their paper. 
Our discovery of the the east-west outflow strongly support their suggestion. 
Moreover, class II \meth\ masers are exclusively associated with massive young stars \citep{Hill2005}, therefore, M1 provides additional evidence that P1-SMA1 is forming massive star(s).

\begin{figure*}
\centering
\includegraphics[width=0.7\textwidth,angle=0]{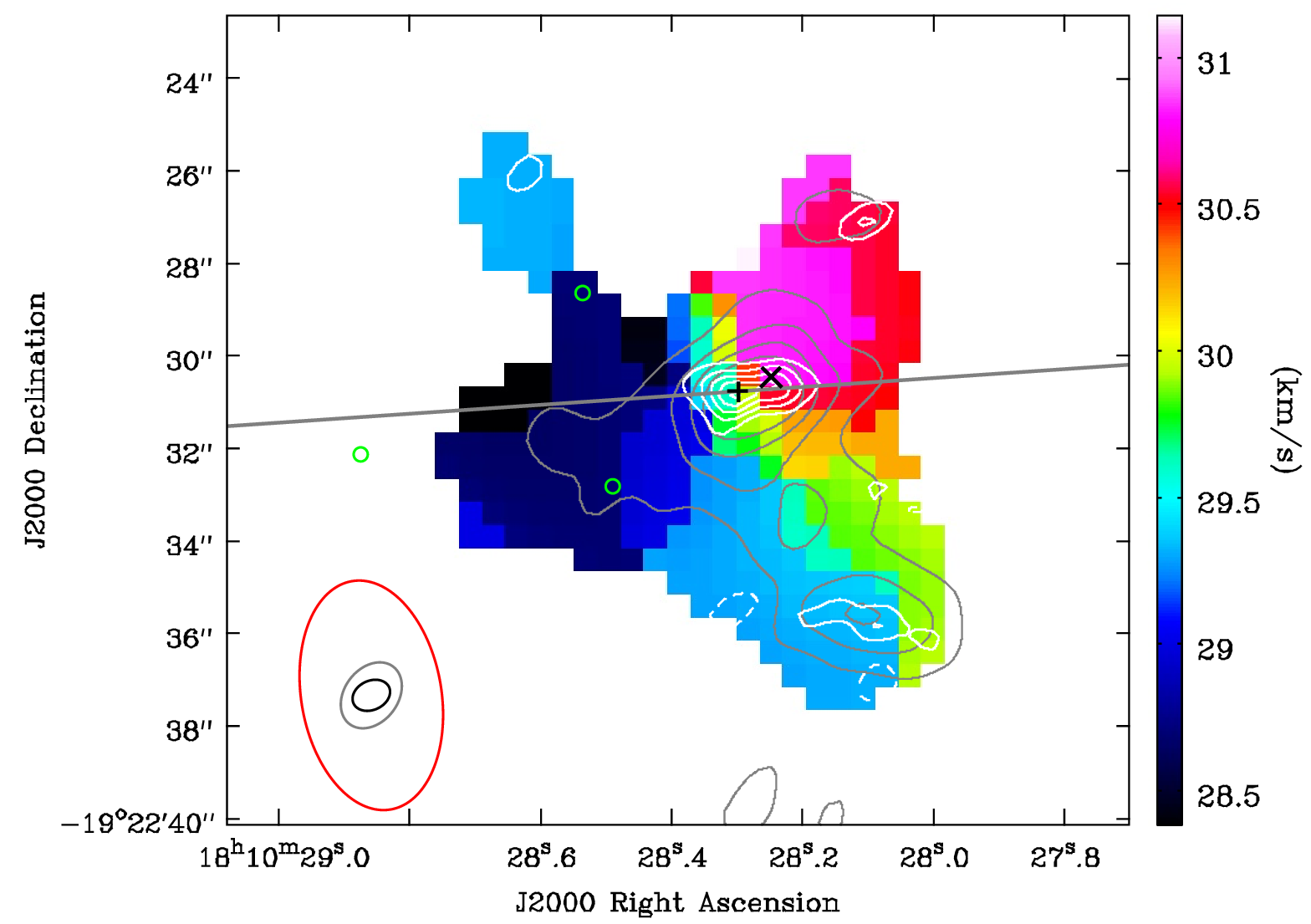}
\caption{
Flux weighted velocity field of \nh3 \22.
The grey contours show the SMA 880\um\ continuum image as in Fig. \ref{fig:con}(c), but contoured by 20\% of the peak (76\mjy) for clarity.
The white contours show the highest resolution SMA 880\um\ image made from the Extended configuration only; the contours are $\pm(3,6,9,...)\sigma$, where $\sigma = 1.7$\mjy beam$^{-1}$.
Labelled symbols denote \water\ maser W2 ($+$), class II \meth\ maser M1 ($\times$), and the three 2MASS point sources ($\circ$).
The ellipses in the bottom left corner represent synthesized beams for the \nh3 image and the two SMA 880\um\ images, from the largest to the smallest respectively.
The thick grey line denotes the SiO outflow as in Fig. \ref{fig:p1.outflow}.
}
\label{fig:p1.nh3_22}
\end{figure*}

From the morphology of high velocity SiO emission we locate the outflow driving source to be within the dust condensation P1-SMA1. However, the large scale SiO emission cannot pinpoint the driving source, because P1-SMA1 is closely associated with multiple sources: the \meth\ maser M1, the \water\ maser W2, the \her\ point source \#9, and the SMA dust condensation. The \her\ and \spt\ images however do not have sufficient resolution to distinguish dust structures associated with M1 and W2. To reach the highest possible resolution with SMA, we make an 880\um\ image using only the data from the extended configuration, and obtain a $0''.8 \times 0''.6$ ($3000 \times 2000$ AU) synthesis beam. The image is plotted with white contours in Fig. \ref{fig:p1.nh3_22}, in comparison to the masers, the outflow axis, and large scale velocity field from \nh3 \22. Surprisingly, P1-SMA1 contains an elongated dust structure, and the major axis of the structure matches the axis of the outflow. The elongated structure has an integrated flux of 76\mjy\ (5\msun\ assuming 25 K dust temperature), and is composed of at least two substructures. The eastern feature (3\msun) coincides with W2, and the western feature (2\msun) near M1. Plausibly, these two dust features indicate a proto-binary system with a separation of 2500 AU. High-mass stars are known to be born mostly as twins \citep{Chini2011_twins,Chini2013_multiplicity}, and a recent ALMA discovery also suggests a disc associated with a proto-binary system in G35.20-0.74N \citep{Sanchez2013_G35.2disk}, suggesting the multiplicity can be traced back to the embedded phase. It is also possible that one of the dust features is heated by the outflow driven by a protostar embedded in another dust feature.

We discuss two possibilities on the location of the outflow driving source.
Possibility I: the driving source is located within the disc traced by the \meth\ maser spots, therefore coincides with M1. This supports that \meth\ masers trace disc other than outflow \citep{Pestalozzi2004_meth}. W2 would locate in the blue lobe of the outflow, but its velocity is red-shifted (37.5 \kms; \S\,\ref{sec:maser}). 
Possibility II: the driving source coincides with W2 and the more massive eastern dust feature. The \nh3 velocity field seems to support this scenario since W2 resides closer than M1 to the transition from blue- to red-shifted velocity (Fig. \ref{fig:p1.nh3_22}). In this scenario, M1 traces the red lobe of the outflow \citep{DeBuizer2003_meth,Kurtz2004_meth}, but its velocity range spans blue to red (22--34 \kms; \S\,\ref{sec:maser}), and it becomes difficult to explain the ordered velocity field of M1 that could otherwise perfectly trace a disc \citep{pillai06b}.
Among all the pieces of evidence, the strongest evidence seems to be the maser spots associated with M1. We therefore regard the possibility I is more likely compared to possibility II. Deep near to far-infrared imaging with higher resolution is needed to pinpoint the driving source, and thereby help further disentangle the controversy about class II methanol masers tracing disc \citep{Pestalozzi2004_meth} or shocked gas \citep{DeBuizer2003_meth,Kurtz2004_meth}.
As far as we see with current data, our results combined with \cite{pillai06b} supports the disc scenario.

\subsection{Evolutionary Sequence} \label{sec:evo.clumps}

In \S\,\ref{sec:chem} we have seen that the chemical differentiation in cores suggests an evolutionary sequence of the cores.
On larger scales, the 1 pc clumps in the Snake appear to be at different evolutionary stages as well. P1 and P6 are the most evolved clumps, where P1 is slightly more advanced than P6. Other clumps appear to be younger. For example, \cite{Tackenberg2012} identified five starless clumps within the Snake nebula.
The brightness temperature of the detected lines in P1 and P6 are typically 1--5 K (Fig. \ref{fig:allspec}). If \gb\ were located at the distances of \gapa\ or \gcca, the brightness would be 2 or 3 times lower even if the structures are unresolved, whereas in reality the structures within clumps are well resolved by SMA. The sensitivities of the SMA observations \citep{qz09,me11,qz11} were enough to detect such lines. The fact that previous SMA observations did not detect lines except CO in \gapa\ and \gcca\ suggest that those clumps are chemically younger than \gbpa\ and \gbpb.

Thanks to our coordinated SMA and VLA observations, we are now able to compare the relative evolutionary stages of the five aforementioned IRDC clumps. The basis of this comparison is that the clumps are of similar mass and size ($\sim 10^3$\msun, $\sim 1$ pc), and that they evolve along the same path. Table \ref{tab:clumps} lists the observed properties of the clumps, including global properties (mass and luminosity), and star formation signatures (dense molecular lines, outflows, dense cores, cylindrical collapse, and maser detection). Among all the clumps, only \gbpa\ shows all the star formation signatures and only \gcca\ shows none of the signatures. Thus, clump \gcca\ is the youngest and \gbpa\ is the most evolved clump, while other clumps are in between. This relative evolution is consistent with the global luminosity-to-mass ratio (Table \ref{tab:clumps}). Clumps with a similar mass but at a later evolutionary stage have a higher luminosity due to increased protostellar activity, resulting a higher luminosity-to-mass ratio \citep{Sridharan2002_HMPOs,Molinari2008_MYSO_evo,rath10}.

Our sensitive high-resolution observations offer a first view of how chemical and physical properties may evolve over time at various spatial scales relevant to the hierarchical fragmentation. Evolution is found at scales from 1 pc clumps, to 0.1 pc cores, and down to 0.01 pc condensations.
The reason behind this may be due to an inhomogenous initial condition or a somewhat competitive mechanism amongst these star formation seeds.
The spectra of these ``star formation seeds'', like the ones presented in Fig. \ref{fig:allspec} provide an excellent testbed for chemical models. In particular, comparing spectra from one region avoids systematic and calibration uncertainties, offering a powerful tool for assessing how the seeds grow chemically.

\begin{table*}
\begin{minipage}{160mm}
\caption{Clump Properties $^a$ \label{tab:clumps}}
\begin{tabular}{lccc ccc ccc cl}
\hline
{Clump} &
{$d$ $^b$} &
{$M$} &
{$L$} &
{Line $^c$} &
{Outflow} &
{Dense} &
{Cylinder} &
{\water} &
{\meth} &
{\meth} &
{Ref.$^d$} \\
{} &
{(kpc)} &
{($10^2$\msun)} &
{($10^2$\lsun)} &
{} &
{} &
{core} &
{collapse} &
{maser} &
{class I} &
{class II} &
{}
\\
\hline
\gcca\   &6.5    &18	&$\ll$4.6	&$-$  &$-$  &$-$  &$-$  &$-$  &$-$  &$\cdots$	&1,2 \\
\gccb\   &7.3    &12	&4.6	&$+$  &$-$  &$+$  &$+$  &$+$  &$-$  &$\cdots$	&1,2 \\
\gbpb\   &3.6    &9.3	&1.5	&$+$  &$+$  &$+$  &$+$  &$-$  &$-$  &$\cdots$	&3,*    \\
\gapa\   &4.8    &10	&1-21	&$-$  &$+$  &$+$  &$+$  &$+$  &$-$  &$\cdots$	&4,5,6 \\
\gbpa\   &3.6    &12	&12-14	&$+$  &$+$  &$+$  &$+$  &$+$  &$+$  &$+$		&3,7,* \\
\hline
\end{tabular}

\medskip
\textbf{Note}:
\\$^a${ $+$ means ``yes''; $-$ means ``no''; $\cdots$ means no data.}
\\$^b${ Kinematic distance.}
\\$^c${ Detection of lines other than CO isotopologues.}
\\$^d${ References for mass and luminosity:
1. \cite{Swift2009}, 
2. \cite{qz11},
3. \cite{Ragan2012_EPoS}, 
4. \cite{qz09},
5. \cite{wy08},
6. \cite{rath10},
7. \cite{pillai06b},
{*: this work.}
}
\end{minipage}
\end{table*}

\section{Conclusions} \label{sec:sum}

We study fragmentation of two massive ($\gsim 10^3$\msun), low-luminosity ($\lsim 10^3$\lsun), and dense ($\sim 8 \times 10^4$ \cm3) molecular clumps P1 and P6, the most likely sites of high-mass star formation in IRDC \gb, using high resolution SMA and VLA observations. The achieved mass sensitivity is better than the Jeans mass at the clump scale, and our main findings are as follows.

(1) High-resolution, high-sensitivity SMA continuum images at 1.3 and 0.88 mm resolve two levels of fragmentation in both P1 and P6: the clump fragments into 6 dense cores, some of which further fragment into even smaller condensations. While the clump fragmentation is consistent with a cylindrical collapse, the masses of the dust cores and condensations are much larger than the respective thermal Jeans masses. This is similar to what was found in IRDC clumps \gapa\ and \gccb, suggesting that turbulence supported fragmentation is common in the initial stages of massive star formation.

(2) Molecular outflows, masers, shocked \nh3\ emission, as well as hot core lines and mid-IR point sources all indicate active ongoing star formation in cores P1-SMA1,2 and P6-SMA6,2,5. The discovery of an east-west outflow associated with P1-SMA1, together with previous studies, points to an outflow-disc system. A close-up view of the system further suggests a possible proto-binary system, which deserves further study with better resolution.

(3) Enrichment of ortho-\nh3 is found associated with all the three identified outflows in clump P6, and the enrichment tends to increases along the outflow downstream. The derived ortho/para {abundance} ratios are  $1.1\pm0.4$, $2.0\pm0.4$, and $3.0\pm0.7$, {among} the largest OPRs ever observed in Galactic star formation regions while less than that in starburst galaxies, although observations of even higher transitions are needed to confirm the high ratios.

(4) Chemical differentiation between cores suggests variations in evolutionary stages among the cores. This effect is also seen at smaller scales down to condensations, and to larger scales up to clumps. Accordingly, an evolutionary sequence is inferred for cores and clumps, respectively.

This study is part of our coordinated SMA and VLA observing campaign to study the earliest stages of massive star formation \citep[][and this work]{qz09,qz11,me11,me12,me13_ppvi}. Our observations reveal a comprehensive view of the early stages prior to the hot core phase, from the quiescent clump \gcca\ to clumps that harbor accreting protostars like \gbpa.
The overall result shows that:
(a) turbulence is more important than thermal pressure in the initial fragmentation which leads to subsequent clustered star formation;
(b) the hierarchical fragmentation leads to seeds of star formation which grow inhomogeneously, as well as in a hierarchical way.

\section*{Acknowledgements}
{We thank the referee Michael Burton for valuable comments that helped clarify the manuscript.}
K.W. acknowledges support from the ESO fellowship,
the European Union Lotus post-doctoral fellowship under the frame of Erasmus Mundus Action during his stay at Kapteyn, and the SMA pre-doctoral fellowship and China Scholarship Council during his stay at Harvard CfA. Y.W. and K.W. thank the support by China Ministry of Science and Technology under State Key Development Program for Basic Research (2012CB821800).
{Q.Z. acknowledge the partially support by the NSFC grant 11328301.
S.E.R. is supported by grant RA 2158/1-1, which is part of the Deutsche Forschungsgemeinschaft priority program 1573 (``Physics of the Interstellar Medium'').}
This research made use of open-source Python packages \textsc{APLpy} and \textsc{PySpecKit} \citep{APLpy,pyspeckit}.


\end{document}